\documentstyle[12pt,amssym,aasms4]{article}
\lefthead{Gardini et al.}
\righthead{Cluster mass function in mixed models}

\def\be{\begin{equation}}
\def\ee{\end{equation}}
\def\bea{\begin{eqnarray}}
\def\eea{\end{eqnarray}}
\def\mincir{\raise -2.truept\hbox{\rlap{\hbox{$\sim$}}\raise5.truept
\hbox{$<$}\ }}
\def\magcir{\raise -4.truept\hbox{\rlap{\hbox{$\sim$}}\raise5.truept
\hbox{$>$}\ }}

\begin{document}
\title{CLUSTER MASS FUNCTION IN MIXED MODELS}
\author{A. Gardini, S.A. Bonometto, G. Murante}
\affil{Dipartimento di Fisica G. Occhialini -- Universit\`a 
di Milano--Bicocca\\
 \& \\
INFN sezione di Milano -- Via Celoria 16, I20133 Milano, ITALY \\
e-mail: Alessandro.Gardini@mi.infn.it
}
\authoremail{Alessandro.Gardini@mi.infn.it}

\begin{abstract}
 
We study the cluster mass function in mixed dark matter (MDM) models, using 
two COBE normalized simulations with $\Omega_h = 0.26$ and $n=1.2$, and 
$\Omega_h = 0.14$ and $n = 1.05$, both with 2 massive $\nu$'s (MDM1 and MDM2, 
respectively). For the sake of comparison, we also simulate a CDM model with 
spectral index $n=0.8$ (TCDM), also COBE normalized. We argue that, in our 
non--hydro simulations, where CDM particles describe both actual CDM and 
baryons, the galaxy distribution traces CDM particles. Therefore, we use them
to define clusters and their velocities to work out cluster masses. 
As CDM particles are more clustered than HDM and therefore have, in average, 
greater velocities, this leads to significant differences from PS predictions. 

Such predictions agree with simulations if both HDM and CDM are used to 
define clusters. If this criterion is adopted, however, we see that: (i) MDM 
corresponds $\delta_c$ values slightly but systematically greater than CDM; 
(ii) such $\delta_c$ exhibit a scale dependence: on scales $\sim 10^{14}
M_\odot$, we find $\delta_c$$\sim 1.7$ or 1.8 for CDM or MDM, respectively; 
at greater scales the required $\delta_c$ decreases and a substantial cluster 
excess is found, at the large mass end ($ M > 10^{15} M_\odot$).

Clusters defined through CDM in MDM models, instead, are less numerous than 
PS estimates, by a factor $\sim 0.3$, at the low mass end; the factor becomes 
$\sim 0.6$--0.8, depending on the mix, on intermediate mass scales 
($\sim 4$--$5\, h^{-1} 10^{14} M_\odot$) and almost vanishes on the high 
mass end. Therefore: (i) MDM models expected to overproduce clusters over 
intermediate scales are viable; (ii) the greater reduction factor at small 
scales agrees with the observational data dependence on the cluster mass $M$
(which, however, may be partially due to sample incompleteness); (iii) the 
higher spectral normalization is felt at large scales, where MDM models 
produce more objects (hence, large clusters) than CDM. MDM1 even exceeds 
Donahue et al. (1998) findings, while MDM2 is consistent with them.

Simulations are performed using a parallel algorithm worked out 
from Couchman AP3M serial code, but allowing for different particle 
masses and used with variable time steps. This allowed to simulate a cubic box 
with side of 360$\, h^{-1}$Mpc, reaching a Plummer resolution of 
40.6$\, h^{-1}$kpc, and using $(3 \times) 180^3$ particles. 

\vskip 0.2 truecm


\end{abstract}


\section{Introduction}

Cosmological N--body simulations have become the basic tool to study 
the non--linear stages of structure formation and evolution in the Universe.
In the least few years, the use of parallel computers has allowed
simulations with an increasingly wide dynamical range 
and it is now possible to simulate boxes, with side up to hundreds Mpc's, 
with distance and mass resolutions well below the galaxy cluster scale.

In this paper we shall report the results of three N--body simulations,
in a box of side 360$\, h^{-1}$Mpc ($h$ is the Hubble parameters in units 
of 100 km${\rm \, s^{-1} Mpc^{-1}}$), aimed to study the properties of galaxy 
clusters. In particular, we shall concentrate on mixed models, which,
up to now, were sometimes disregarded in large parallel simulations.
Two of the models tested are therefore mixed models, both with 2 massive 
neutrino ($\nu$) species; $\nu$ masses are $m_\nu \simeq 3.02\, $eV and 
1.63$\, $eV, to yield $\Omega_h = 0.26$ and 0.14, respectively. 
(More details on the models are given below; see, in particular,
Table I.) Models were selected, first of all, to approach observational data. 
In this work, however, we wish to quantify a peculiar behaviour of
the cluster mass function, that we expect to occur for mixed models.
Within this context, the third model, a tilted CDM, was mostly selected for 
the sake of comparison.

Clusters of galaxies are the largest bound systems in the Universe and are 
also fairly rare objects. They have been systematically studied in the recent 
years, using both optical and X--ray data. In principle, this allows to use 
them to provide stringent constraints to cosmological models.
The behaviour of their mass function, in particular, is a critical test.
Mass function data are given and/or discussed in several recent papers 
(\cite{Eke,Col,Via,Mo,Bor1,Gir,Pos}).

For any cosmological model, the Press $\&$ Schechter (PS)
approach provides a semi--analytical estimate of such mass function 
and model simulations already showed a basic agreement
with such estimates. Some discrepancies were however outlined at the low and
high mass ends, which will be also confirmed by our analysis.

There is however a well defined point that we wish to explore in this work.
We shall show that the presence of a hot 
component may substantially widen the gap between expected and
observable mass functions. The qualitative reasons of this effect 
can be explained soon, but the extent and the mass dependence of the gap
can only be numerically explored.

As is known, fluctuations in the different components of mixed models
evolve differently. On the scales where hot fluctuations were initially 
erased, $\nu$'s can be trapped again later on, as they become slower,
by fluctuations which survived in CDM (and baryons). Eventually, in
most reasonable models, at $z=0$, the transfer functions of the three 
massive components are again almost equal. At the non--linear level, 
instead, $\nu$'s can be distributed differently from CDM and baryons, 
even at $z=0$. Several simulations (see, e.g., 
\cite{Kly2}) showed their final distribution,
which follows the shape of non--linear structures, but with a
tendency to stay away from structure knots, rather inhabiting
peripheral regions. Such behaviour can be more or less pronounced, 
according to the scale considered.

In the real world, baryons are the main mass tracer. Galaxies emit
because of their presence and diffuse gas emits radiation
because of the electrons accompanying them. Locating baryons in a 
simulation with no hydrodynamics may be hard. E.g., in a recent work,
Valdarnini, Ghizzardi $\&$ Bonometto (1999) have shown that using
CDM instead of baryons, to study the global cluster structure,
may be misleading and surely changes the scores of various cosmological models,
when compared with X--ray data. The question exists also for simulations
involving no hot component, but the presence of a hot component might
complicate the answer.

In principle, however, the outcome would be simple. When initial
conditions for the simulation are set at $z_{in}$, we use separate transfer
functions for HDM, CDM and baryons. The latter two components,
however, have almost identical transfer functions, as baryon infall
in potential wells, set by CDM at the end of recombination, is already 
complete. At $z_{in}$, instead, HDM is still differently distributed. Moreover,
$\nu$'s keep average velocities ${\bar v }/c \simeq 5.3 \cdot 10^{-4} 
(1+z_{in}) 4h^2 (m_\nu/{\rm eV})^{-1}$, of thermal origin.
Simulations are then performed with two kinds of particles, conventionally
called CDM and HDM. CDM particles, however, account both for baryons and 
actual CDM. HDM particles, instead, account for $\nu$ behaviour and,
since the beginning of the simulation, have quite a different treatment,
according to their peculiar properties, including their large average
velocity $\bar v$. The obvious conclusion is that, 
in a dissipationless simulation,
we should seek baryons where we put them, $i.e.$ with CDM {\sl particles}.

Previous authors, however, distinguished between galaxy and gas
setting, even in dissipationless simulations. E.g., Kofman et al. (1996),
who performed a PM simulation of MDM with 2 $\nu$'s of mass
$m_\nu = 2.3\, $eV and $n=1$, in a box of $50\, h^{-1}$Mpc ($h=0.5$), using
$3 \times 256^3$ particles and a force resolution $\sim 190\, h^{-1}$kpc
(hence a greater mass resolution and a slightly worse force resolution
in a much smaller box, if compared with simulations described in this work),
stated that distribution and motion of galaxies are primarily
signatures of CDM particles. However, to try to spot X--ray emitting gas,
they set it in hydrostatic equilibrium, assuming a spherically
symmetric static gravitational potential given by a King--like profile,
while the temperature profile was worked out using data on the X--ray
surface brightness of the Abell cluster A2256, which is a rich, spherically
symmetric, almost relaxed cluster, quite similar to Coma. 
This kind of prescriptions are however almost useless if simulations
are compared with galaxy data; CDM and galaxy distributions can be
assumed to be similar on the scales resolved in this work.

Therefore: (i) Even in mixed models, to compare simulations with data, 
we define clusters using CDM particles only. (ii) In order to evaluate cluster
masses to fit data, we estimate them through the velocity
distribution of CDM particles only. Observational 
estimates, of course, are performed along a complex pattern.
We shall be using recent results provided by Girardi et al. (1998),
obtained using ENACS data, and previous results of the same
group (Biviano et al. 1993). In principle we should work out
a mock catalog from our simulation and use it, so to reproduce
all biases that data may have and use all prescriptions followed to try to
overcome them. We plan to do so in a forthcoming work. Meanwhile, however, 
it is clear that cluster masses are to be determined through velocity
distributions. The procedure to work out such {\sl virial} masses, 
aiming to approach what is done in observations, 
will be better described below.

Week gravitational lensing, instead, depends on the whole mass,
carried either by CDM or HDM particles, through the potential it produces.
We shall make no comparison here with masses obtained through lensing data.
Just notice that, according to our arguments,
in mixed models, they may be grater than {\sl virial} masses by 
a factor which exceeds unity by more than $\Omega_h/\Omega_c$.

If we associate these points
to the fact that CDM and HDM are distributed differently and expected
to have different velocity distributions, it is clear that significant
discrepancies from PS expectations can be present in mixed models.
Let us examine which could be their trend.

First of all, CDM and HDM distributions are more and more alike
as we go to greater scales. Henceforth, above a suitable scale,
the peculiarity of mixed models might become negligible. 
It is however hard to give an analytical estimate of such scale,
if it exists. Then, PS mass functions are worked out from transfer functions.
If they are equal for the three components at low $z$, we can expect
that PS predictions hold if clusters are defined using both 
cold and hot particles. This expectation is however to be numerically verified.
Assuming that this is the case, the smoother distribution expected
for $\nu$'s, in respect to CDM, may lead to a decrease
of the amplitude of the actual cluster mass function. In particular,
we shall see some examples where $\nu$ clouds around smaller
CDM condensations create bridges among them. In similar cases, clusters 
defined using all particles turn into multiple systems of smaller objects, 
when only CDM particles are considered.
But, quite in general, HDM is mostly periferal and helps to
extend the volume where CDM particles are attributed to the cluster.
Forgetting HDM particles, therefore, leads to a decrease of the very
CDM particles belonging to most clusters. This causes a decrease
of cluster masses which may exceed the decrease due to neglecting
HDM. Its extent, however, is to be numerically tested.

However, besides using CDM particles only, we must recall that real mass 
measurements are not based on counts of particle {\sl numbers}; obviously, this
can be done only in simulations. Optical and $X$--ray data, 
instead, are essentially obtained
through velocities. If HDM particles tend to stay farther 
from the main condensations, preferring regions where
the potential well is shallower, we can expect that CDM particles,
in average, take an extra share of the overall kinetic energy, that
will depend on detailed $\nu$ parameters.
If their kinetic energy is used to test the cluster mass, we can
expect a systematic overestimate. Once again, a quantitative 
evaluation of the effect can hardly be obtained without numerical means.

These are the main physical points that we wish to explore through our 
simulations. As we shall see, such effects do exist and are
quantitatively relevant.

Exploring the world of mixed models can be particularly expensive, from the 
numerical point of view. In a given standard CDM (sCDM) simulation,
a change of normalization can be interpreted as a change of the time when
$z=0$; furthermore, sCDM can be turned into $\tau$CDM (see, e.g., 
\cite{Bon,Whi,Mcn,Bha,Han}) by rescaling the box and the shape parameter 
$\Gamma$, which, in the transfer function of $\tau$CDM models, always
appears in the combination $\Gamma/k$. (See section 3 for a discussion
on the meaning of $\Gamma $.) Such multiple interpretations of a single 
simulation do not exist for mixed models. Even apart of the $\tau$CDM variant, 
a change of normalization should be accompanied by a shift of the velocities 
of hot particles at the simulation start. 

Besides the restriction ``one model--one simulation'', mixed models
need at least 3 particles instead of 1, to obtain the same level of
resolution obtainable with pure CDM. In fact CDM and HDM components 
need different particles and the latter ones must be double in number,
to account for thermal velocities, without introducing
a spurious non--zero linear momentum density.

Further families of models which can be simulated using one kind of 
particles only are $\Lambda$CDM and models with total density 
parameter $\Omega_o < 1$ (OCDM). The importance of $\Lambda$CDM models has 
greatly increased after an improved analyses of SNIa data (\cite{Per,Rie})
indicated an accelerating cosmic expansion. OCDM models, instead, seem 
favoured by the presence of high redshift large scale matter condensations 
(\cite{Bah,Don}) and by the statistics of arclets
(\cite{Bar}). The latter conclusions, however, might have to be partially
reconsidered on the light of some results of this analysis.

This work was made possible by our implementation of a parallel N--body 
code, based upon the serial public AP3M code of Couchman (1991), extended
in order to treat variable mass particle sets and used varying the time--steps,
when needed. Our 
simulation, dealing with a 360$\, h^{-1}$Mpc cubic box, needed either 180$^3$ 
(for pure CDM) or $3 \times 180^3$ (for MDM) particles. The force resolution
is set by a Plummer--equivalent smoothing parameter $\epsilon_{pl} \simeq
40.6\, h^{-1}$kpc. Our comoving force and mass 
resolutions approach the limits of the computational resources of
the machine we used (the HP Exemplar SPP2000 X Class processor
of the CILEA consortium at Segrate--Milan). Such resolutions
are close to the ones of the four simulations for pure CDM
with different initial conditions described by Colberg et al. (1997), 
Thomas et al. (1997), or Cole et al. (1997). 
See also Gross et al. (1998).
More details on the simulations are given on section 3.

The plan of the paper is as follows. In the next section we discuss 
first the linear features of the models treated. In section 3 we give
all technical details concerning our simulations. Their outputs are 
then used to show the evolution of the spectra; the final spectrum, at $z=0$, 
will then be compared with APM data and current approximated expressions, 
finding discrepancies which characterize mixed models (section 4).
In section 5, five different algorithms, used to select clusters, are briefly
discussed. An extended comparison among their outputs is outside
the scope of this work; our basic issues are contained in Section 6,
where the effects of the presence of a hot component,
on the cluster mass function, will be suitably summarized.
Section 7 is devoted to a final discussion.

\section{Linear features of the models}

Quite in general, to define a model we require: background metric, substance 
mix and primeval spectrum. The background metric is fixed by the Hubble 
parameter $H = 100\, h\, {\rm km\, s^{-1} Mpc^{-1}}$ and the overall density 
parameter $\Omega_o = \rho_o /\rho_{cr}$ ($\rho_o $:
present average density, $\rho_{cr} = 3 H^2 / 8\pi G$:
critical density). Substance is fixed by partial density parameters. 
E.g., $\Omega_b = \rho_b/\rho_{cr}$ or $\Omega_c = \rho_c/\rho_{cr}$, which are
the ratios between baryon or cold--dark--matter (CDM) and critical densities.
Hot--dark--matter (HDM) is fixed by two of input data: besides of
$\Omega_h$, we must give its number of spin degrees of freedom $g_{\nu,m}$.
In this paper, instead, $\Omega_\Lambda \equiv 0$. 
Further quantities to be specified are the present CBR temperature and the 
effective number of massless--$\nu$ spin degrees of freedom ($g_{\nu,0}$). 
Finally, early deviations from
homogeneity are parametrized by the amplitude $A_\Psi$ and 
the spectral index $n$ of the initial fluctuation spectrum 
$$
P(k) = {2\pi^3 \over 3} {A_\Psi \over x_o^3} (x_o k)^n
\eqno (2.1)
$$ 
(here: $x_o$ is the comoving horizon distance;
$k = 2\pi/L$ is the wave--number related to the comoving length
scale $L$ and mass scale $M=(4\pi/3) { \rho_o} L^3$).

In this section, we report predictions on 
large scale structure (LSS) obtainable from
the linear theory, for the models we simulated. In all our models
$\Omega_o = 1$ and the Hubble parameter $h=0.5$.
The other model parameters are shown in Table I, together with the above
mentioned predictions, worked out from transfer functions.

\placetable{tab1}

Transfer functions are obtained by solving numerically a large set
of coupled differential equations. The set is particularly wide for
mixed models (see, e.g., \cite{Bono2,Bono3,Ach,Hol,Ma2}). 
Recent work \cite{liddle96,prima95,smith98,gross98,gawi98,Pie,Bono1}
has shown that a wide parameter space exists, for which
mixed models are consistent with LSS features that we can estimate within
linear theory.

Assuming the presence of scalar fluctuation modes only, in $\Omega_o = 1$ 
models, the angular fluctuation spectrum of CBR, for small $l$ values,
can be approximated with its Sachs \& Wolfe expression
(see, $e.g.$, \cite{Ma2,Sel}):
$$
C_l \simeq (2\pi/3)^2 A_{\Psi} \int_0^{\infty} 
(dk/k) (x_o k)^{n-1} j_l^2 (x_o k) ~,
\eqno (2.2)
$$
($j_l$ are Bessel fuctions). As
$$
C_2 = {4\pi \over 5}\left(Q_{rms, PS} \over T_o \right)^2
\eqno (2.3)
$$
(see, $e.g.$, \cite{Ma2,Sel})
the spectral amplitude $A_\Psi$ is obtainable from data for $Q_{rms, PS}$.

The models we simulate have different $n$ values: for the tilted CDM
(TCDM) we took $n = 0.8$ while, for MDM1 and MDM2, $n=1.2$ and $n=1.05$,
respectively. Among physical unknowns, there is the contribution of primeval
gravitational waves to the observed $Q_{rms, PS}$ value. As a working
solution, in order to leave room for such signals, without spoiling the
fit with COBE data, we decided to work out $A_\Psi$ values from lower
3$\, \sigma$ limits on $Q_{rms, PS}$. 

To our knowledge, simulations of mixed models with $n > 1$
were never performed before, for any box size and dynamical range,
except by Lucchin et al. (1996), who however considered a model with
$\Omega_h = 0.3$ and a single $\nu$ family, normalized to the
central COBE value, and therefore significantly violating some 
linear constraints (e.g., $N_{cl}$, see below).

Using the transfer function $T(k)$, we can evaluate the mass variance over the 
comoving scale $L$:
$$
\sigma^2 (L) = {\pi \over 9} \left(x_o \over L \right)^{n+3} A_\Psi
\int_0^\infty du u^{n+2} T^2 \left(u\over L\right) W^2 (u)~.
\eqno (2.4)
$$
Here, for a top--hat window function, $W(u) = 3(\sin u - u \cos u)/u^2$.
For $Lh = 8$ and 25 Mpc, we evaluate $\sigma_8$ and $\sigma_{25}$,
which allow to work out 
$$
\Gamma = 7.13 \cdot 10^{-3} (\sigma_{ 8}/\sigma_{25})^{10/3}~;
\eqno (2.5)
$$
the connection between this general definition and some current approximated
expressions, like $\Gamma \simeq \Omega_o h$, can be found in \cite{ebw92}.
Its values, as well as the values of $\sigma_8$, are
reported in Table I. Peacock \& Dodds (1994),
using APM data, and Borgani et al. (1994) using Abell/ACO samples, obtained 
the (2$\, \sigma$) intervals 0.19--0.27 and 0.18--0.25 for $\Gamma$.

According to the PS approach, the expected cumulative number density of 
clusters is related to $\sigma_8$ and $\Gamma$ and reads:
$$
n(>M) = \sqrt{2/\pi} (\rho/M) \int_{\delta_c/\sigma_M}^\infty 
du [M/M(u)] \exp(-u^2/2) ~.
\eqno (2.6)
$$
Here $M(u)$ is defined so that the mass variance (expressed in function
of mass--scale, instead of length--scale) $\sigma_{M(u)} = \delta_c/u$; 
$\delta_c$ values from 1.4 to 1.8 (\cite{Pee}) were considered.
Making use of eq. (2.5) with a top--hat window function, the transfer
function can be used to compute $N_{cl} = n(>M) \times R^3$ for 
$R = 100\, h^{-1}$Mpc and $M = 4.2\, 10^{14} h^{-1} M_\odot$. 
Values of $N_{cl}$, obtained within the PS approach, 
are given in Table I, for different models.
An observational range is $N_{cl} = 4-6$ (\cite{WEF,Biv,Eke,Gir}),
but this result is still subject to a number of uncertainties
and values within a factor 2 from upper and lower limits cannot be
safely rejected. However, even with such extra freedom, our MDM1 model
would appear out of the observational range.

We must however bear in mind that PS estimates are based on the
transfer function at $z=0$, which are almost identical for all
dark matter and baryon components. As we shall see, non linear
effects lead to greater condensations in CDM (and baryons)
than in HDM, at all $z$'s. Testing the consequences of such effects is
one of the aims of this work. Taking them into account, for reasons
widely discussed below, $N_{cl}$ may become substantially smaller
and even MDM1 will be found compatible with data.

Further severe tests for tilted or mixed models concern the formation of 
structures at high $z$. Among them, one of the most stringent is the amount 
of gas expected in Damped Lyman--$\alpha$ systems, which can be expressed 
through the parameter $\Omega_{\rm gas} = \alpha \Omega_b \Omega_{\rm coll}$.
Here $\alpha$ is an efficiency parameter ($\mincir 1$).
Let then $\sigma_M(z) $ be the mass variance, worked out from  the 
transfer function ${\it T}_k(z)$, for the mass--scale
$M$ at redshift $z$. Using its value,
$$
\Omega_{\rm coll} = {\rm erfc}[\delta_c/\sqrt{2} \sigma_M(z)]~.
\eqno (2.7)
$$
is easily obtained.
Accordingly, we evaluate $L_\alpha \equiv \Omega_{\rm gas} 
\times 10^{3}  /\alpha $, taking $z = 4.25$ and $M = 5\cdot 10^{9} h^{-1}
M_\odot$ in $\sigma_M(z)$, 
$\delta_c = 1.55$ and a top--hat window function (see, however, 
\cite{Ma1,Kly1}).
Data provided by Storrie--Lombardi et al. (1995)
give $L_\alpha > 2.2 \pm 0.6$, while $L_\alpha $ values expected
for the different models are given in Table I. 

Expected bulk velocities were also evaluated for our models and found 
consistent with POTENT reconstructions of velocity fields. Altogether, 
therefore, our models can be said to predict linear features consistent
with most current linear observational constraints.

\section{The simulations}

The simulations start from the redshift $z_{in}=10$.
The particle sampling of the density field is obtained applying 
the Zel'dovich approximation (\cite{Zel,Dor})
starting from a regular grid. For MDM models, the prescriptions 
of Jing $\&$ Fang (1994) or Borgani et al. (1997) were followed
(see also the seminal paper by \cite{Kly1}) 
We adopt the same random phases in all simulated models.
Particular care is taken to treat the HDM
component, sampled by couples of particles initially
set in identical positions, with opposite thermal velocities, 
in order not to have a spurious linear momentum.
Henceforth, the simulation will contain a double number of
HDM particles compared to CDM. 
In the definition of initial mode amplitudes, the same random numbers
were adopted for cold and hot components. This allows a fair fit
of numerical and analitical amplitude growings, at the initial stages.

As already mentioned, our simulations
study a periodic cubic box of side $L = 360\, h^{-1}$Mpc.
CDM+baryons are represented by $180^3$ particles, 
whose individual mass is $2.22 \cdot 10^{12} h^{-1} M_\odot$ for TCDM,
$1.64 \cdot 10^{12} h^{-1} M_\odot$ for MDM1 and
$1.91 \cdot 10^{12} h^{-1} M_\odot$ for MDM2;
the mass of the $2 \times 180^3$ HDM particles is
$2.89 \cdot 10^{11} h^{-1} M_\odot$ for MDM1 and
$1.56 \cdot 10^{11} h^{-1} M_\odot$ for MDM2.
Outputs of the simulations are preserved at
20 intermediate redshifts $z_i$ set at constant time intervals
$\delta t \simeq 3.26 h^{-1}\cdot 10^8$yr.
The comoving force resolution is given by
the softening length, $\eta \simeq 112\, h^{-1}$kpc. The force $F$ is evaluated
considering each particle as a smoothed distribution of mass, with shape
$ \rho(r) = (48/\pi \eta^4)(\eta/2-r)$ for $r<\eta/2$ (this is the so--called 
S2 shape, \cite{HE}). It behaves as $F \propto 1/r^2$
when $r \ge \eta$. Since the softening of the force is usually referred
to a Plummer shape, $F \propto r/(r^2+\epsilon^2)^{3/2}$, we used a least 
$\chi^2$ test to estabilish the best approximation between the forces
generated by the two different shapes. The minimum $\chi^2$
occurs when $\eta=2.768\epsilon$. In our case, this corresponds to a Plummer
equivalent softening $\epsilon_{pl} \simeq 40.6 h^{-1}$kpc; we will use this
latter value of the softening as our nominal force resolution.

For the sake of comparison, the numerical simulations of
Colberg et al. (1997), Thomas et al. (1998)
have $256^3 \simeq 1.67 \times 10^7$ or $200^3 = 8 \times 10^6$ particles
in a 239.5$\, h^{-1}$Mpc cubic box with $\epsilon_{pl} \simeq 36\, h^{-1}$kpc,
and Cole et al. (1997) have a cubic box of side $\sim 350\, $Mpc,
a softening $\epsilon_{pl} \simeq 90 h^{-1}$kpc with $192^3$ particles.
Gross et al. (1998) 
using a parallelized standard PM code
simulated a mixed model and various CDM variants
in a box containing $(3\times) 384^3$ particles
and $1152^3$ grid points.
They performed simulations both at high and low resolution 
in boxes $300^3$ and $75^3\, h^{-1}$Mpc wide
which correspond grid sizes of $390$ and $65 h^{-1}$kpc.
The claimed softening is $1.5$ times the grid size.

The program used to compute the evolution of 
particles, under the action of gravitational forces, 
is a parallel code developed starting from the public AP3M serial code of 
Couchman (1991). Such code divides the interparticle forces
in long range and short range ones. The long range forces are accounted for 
by the PM ({\it particle mesh}) part, 
which solves the equations of gravitation in the 
Fourier space, where they are essentially algebric, making use of FFT 
to transfer results from coordinate to momentum space and {\sl viceversa}.
The resolution of PM calculations is set by the number of grid cells, 
that we fixed at $256^3$. Further resolution, in our case down to 0.08$\, 
L/256$, is attained by direct summation between neighbour particles 
in the PP ({\it particle particle}) part.
However, where the particle density attains or exceeds $\sim 30$
times the mean value, the evaluation of the short range forces
is performed by repeating the above scheme (refinement),
therefore furtherly subdividing the forces in long and short range ones, and
treating them with a PM and PP calculation respectively. 
The boundary conditions for the refinements are therefore not periodic. 
If the refinement volumes still contain high--density regions, 
the refinement process can continue to deeper levels.

The parallelization of the code was made taking into account the
technical characteristics of the HP Exemplar SPP2000 X Class processor
of the CILEA consortium (Segrate--Milan) we employed. 
The machine architecture is shared--memory. This architecture allows 
to parallelize the calculation without explicitly managing the data 
distribution. 
The SPP2000 of CILEA is composed by two hypernodes, each with
16 HP PA--RISC 8000 processors and a total memory of 4$\, $Gbyte.
The code worked on a hypernode of the machine, using 8 of its processors.
This is due to system--specific requirements.
The memory resources we used were up to $\simeq 1.\, $ and 
$\simeq 1.34 $Gbyte (out of $2\,$Gbyte of shared memory available)
for TCDM and MDM, respectively.
Approximately half required memory is spent to store particle coordinates, 
momenta and masses. The rest of RAM is then needed for all other calculation 
purposes (density field storage, FFT overhead storage, etc.).

The main effort to parallelize the program was concentrated
on PP, both at level 0 and refinements. 
Once the PM part is executed, PP calculations at level 0 
are shared among available processors, as
the linked--list mechanism of the code automatically 
divides consequently the different volumes
where the PP calculation is being performed. 
When the threshold for refinement is attained,
the related operations are executed by a single processor. Further refinements 
are assigned to another processor, etc.; this allows to execute, with the 
setup we used, up to 8 refinements simultaneously. 

As is usual, let us define the speed--up $S_p$
as the ratio of the CPU time used for serial execution versus
the solar time spent for parallel execution on $p$ processors.
Our program speed--up is 
$S_8 \simeq 7.1$ for the PP part at the first steps, but, as is expected,
$S_8 $ slows down to $\simeq 6.4$ at the last steps, due to unbalancing.
As far as refinements are concerned, we reach a value $S_8 \simeq 6.95$
at the last steps. The PM part, instead, has no substantial advantage 
from using system parallel FFT. Altogether, the speedup
remains roughly constant along the simulations,
because of the increasing computational weight
of the short range calculus, as clustering develops and,
in average, never overtakes a value $\sim 3$.

The number of steps made was different for the 3 simulations. 
1000 equal $p$--time steps were used for TCDM (the time parameter 
$p \propto a^{2/3}$; at these late redshifts, the expansion is
nearly matter--dominated, also in the presence of the HDM component, and
equal $p$ intervals yield equal time intervals). MDM1, instead, was run
splitting each of the first 100 above steps into two equal parts. Hence it
used 1100 time steps. The time step choice for MDM2 was more
complex. With reference to the 1000 equal time steps (as used for TCDM), 
we give a succession of number pairs; the former one indicates the step 
numbers out of 1000, the latter one the step numbers they have became
(in order to fulfill the criteria outlined herebelow): [2,10] -- [23,92] 
-- [50,100] -- [125,125] -- [800,400].

Such step choices were dictated by two criteria: (i) Energy conservation.
According to Layzer Irvine equations (see, e.g., \cite{EDFW}), it
had an overall violation $< 3\, \%$ for TCDM and MDM2 and $< 1\%$ for MDM1. 
(ii) Cole et al. (1997) requirements, that the rms displacement of particles in 
a step is $< \eta/4$ and the fastest particle has a displacement $< \eta$
were never violated. 

In fact, for MDM1, the (ii) criterion would have been violated in the first 
100 of 1000 equal time--steps, because of the residual thermal velocity
of $\nu$'s. The problem was even more serious for MDM2. This
led to the subdivision of the initial part of the runs. However, we also 
performed the initial 10$\%$ of MDM1 in 100 time--steps, thus violating
the Cole criterion. Final particle positions are displaced up to
1.6$\, \%$, although the average displacement can hardly be distinguished
from computational noise. (Similar displacements  are found when
the same program is run on two different machines.) 
Accordingly, no discrepancy is present in spectra. We should therefore 
conclude that the (ii) criterion is perhaps too restrictive.

In fig. \ref{fig1} we show 3 slices (one for each simulation),
$ 360\, h^{-1} \times 360\, h^{-1} \times 10\, h^{-1} $Mpc$^3$ wide,
of the simulation volume at $z=0$, projected along their shortest side.
Approximately half of CDM particles are showed. An eye inspection
of the plots shows two clear features: (i) The distributions of
matter, in the two slices, have similar shapes, as is to be expected,
as the two simulations are started from the same random numbers.
(ii) MDM1,2 show more pronounced features on a greater scale, in
respect to TCDM. This should be ascribed to the lack of power
on intermediate scales (100--500$\, h^{-1}$Mpc), which is one
of the characteristics (and problems) of tilted models; the shape of the 
spectra of the two models at $z=0$ can be seen in fig.~\ref{fig3} herebelow.

\placefigure{fig1}

\section{Power spectra and their evolution}

In this section we describe the evolution of the simulation spectra for matter
distributions.

In fig. \ref{fig2} the evolution of the spectrum of the 3 models, starting
from initial conditions and up to $z=0$ is shown. Two sets of 3 plots refer
to the two components (cold and hot) of MDM1 and MDM2 and to their overall
spectrum. The last plot refers to TCDM. 

\placefigure{fig2}

The spectral points are worked out from particle coordinates, performing the 
following operations: On 180 grid points ($\bf n$) we construct a matter 
density field $\rho ({\bf n})$ using the CIC (cloud--in--cells) scheme
(here $\bf n$ is a vector with components $n_i$ indicating the
discrete coordinate of each cell). Let then be $\delta ({\bf n}) =
\rho ({\bf n})/{ \rho_o} - 1$, and let $\hat \delta({\bf k})$ 
be its Fourier transform. Here $\bf k$ values with components
multiple of $k_o = 2\pi / L$, up to 180$\, k_o$, are considered.
The moduli of $\bf k$ span a set of discrete values, to be averaged to
obtain $P(k) = \overline {\vert{\hat \delta({\bf k})} \vert^2 }/k_o^3$
(averaging is over directions and on a modulus interval of width $k_o$).

The plots show that particle distributions are able to reproduce
the initial power spectrum only above a scale $\lambda \sim 10\, h^{-1} $Mpc.
This is a standard feature for simulations,
when initial conditions are set using a grid.
White (1996) suggested an alternative approach
based on a glass (for a recent discussion
of this point and a comparison between initial conditions set
on a grid or on a glass, see \cite{Kne}). 
At variance with PM, P3M simulations explore scales well below
the resolution set by the initial conditions, by exploiting the
particle--particle part of the code. This is a standard and welcome feature
of such simulations and there has been a wide debate on the
reliability of P3M outputs down to such scales 
(see, e.g., Splinter et al. 1998 and references therein). 
Here we keep to the standard approach, but one should bear in mind
that some reserves on this point were risen in the literature. Spectra
are plotted at $z_{in} = 10$, at $z = 3$, $z = 1.13 $ and at $z = 0$.
The two intermediate redshifts correspond to fractions 0.1 and 0.3 of
the total time of non--linear evolution $t_{tot}$.

A further comment concerns the evolution of the spectrum of the hot
component in MDM. Its high--$k$ rise at $0.1\, t_{tot}$ is a well--known
feature, appearing in mixed simulations, and originates from
shot noise, as hot particles cover distances greater than
cell sizes, because of thermal motions. Several attempts are reported,
in the literature, to reduce such effect (an obvious way amounts to
increasing the total number of particles used to simulate the
hot component) and to evaluate how far it can affect final results.
In this work no advanced of sophisticated treatment is applied to
this feature and we follow the standard pattern discussed by
Klypin et al. (1993). Elsewhere, we plan to discuss such effect
in more detail, stressing a peculiar feature, which is to
be overcame and however risks to cause unphysical differences among 
simulations of the same model with different numbers of hot particles, 
unless suitable cautions are taken. In fact, pairs or sets of hot particles, 
initially in the same position, once reaching a distance $\sim \eta$, feel a
mutual gravitational attraction, whose potential energy will cause a decrease 
of their thermal kinetic energy. Such braking effect is due to representing
$\nu$'s through galactic mass particles. In our simulation, we introduced 
an $ad$--$hoc$ option, switching off the interaction within
each hot--particle pair until they reach a mutual distance $\sqrt{3} L/360$,
coinciding with the average initial distance between cold and hot
particles. At $\sim 0.1\, t_{tot}$ the whole option is switched off
and all gravitational interactions are resumed. As is also shown by the
appearence of shot noise at such time, however, the fraction of
particles still in binding danger is then negligible.

In fig. \ref{fig3} we compare the output spectra at $z=0$ with
the linear spectrum at $z=0$, the spectrum at $z=0$ corrected
for non--linearity according to Peacock $\&$ Dodds (1996) and APM 
reconstructed spectral points (\cite{Bau}). 
At variance with fig.~\ref{fig2}, power spectra
here are corrected for the effect of CIC convolution, by dividing them by a 
squared top--hat window--function $W(kR)$, where $R$ is 0.85 times the
particle spacing width. As is shown in more detail by fig. \ref{fig4}, 
in the case of TCDM, simulation outputs almost overlap the 
the spectrum corrected for non--linearity. For MDM, instead, simulation
outputs systematically exceed the theoretical curve, although by a small
amount.

\placefigure{fig3}

\placefigure{fig4}

\section{Cluster mass function}

One of the main aims of the simulations is that of obtaining
a large set of model clusters for each cosmological model, at various
redshifts. This will enable us to study cluster evolution and
to create mock cluster catalogs. It is not clear, instead, how far
the global cluster morphology can be understood within the frame of
non dissipative simulations (see, e.g., \cite{Val}).
Here we shall report some basic results and general 
properties of the clusters selected in the simulation outputs.
In particular, we shall give their cumulative mass function $n(>M)$.

The clusters we consider here were found using a spherical
overdensity (SO) algorithm, yielding the cluster locations, the
radii $R_s$ inside which a density contrast $\delta_{cr} = 180$ is attained
and the total mass $M$ of particles within $R_s$. For mixed models,
the procedure will be applied both to the whole
set of cold and hot particles and to cold particles only,
although only the latter output bears a direct physical significance.

Let us now describe our SO procedure implementation. As a first step,
candidate clusters are located using a standard FoF algorithm, with
linking length $\lambda = \phi \times d$ (here $d$ is the average 
particle--particle separation), giving as outputs groups with more than $N_f$
particles. We then perform the following operations: (i) we find the
center--of--mass $C_M$ of each group and (ii)
we determine the radius $R_g$, inside which the density contrast is 
$\delta_{cr}$ (all particles are to be included, not only those
initially found by FoF). In general, the new center--of--mass is not $C_M$.
The operations (i) and (ii), define a new particle group, on which
the very operations (i) and (ii) can be repeated. The procedure is
iterated until we converge onto a stable particle set. If, at some stage, the 
group comprises less than $N_f$ particles, we discard it. The final $R_g$
is $R_s$. In the actual implementation of SO, it may happen that
the same particle is a potential member of two particle groups. In this case
the procedure assigns it to the more massive one. As a matter of fact,
this has the consequence that, sometimes, more massive groups
swallow smaller ones. A consequence of this choice is a slight decrease
of the total number of clusters, over all mass scales, and this is
confirmed by some preliminary comparisons described below.

In fact, before using this procedure on our simulations, 
we performed a number of checks on test simulations of 
the same cosmological models, with a smaller
mass resolution ($[3 \times] 64^3$ particles, in the same volume).
Clusters found by our SO procedure were compared with those found 
by other standard public cluster
identification algorithms, namely: FoF (\cite{DEFW},
http:\-//www-hpcc.astro.washington.edu/tools/FOF) itself, 
HOP (\cite{Eis}; 
http://www.sns.ias.edu/\char`\~ eisenste/hop/hop.html) 
and DENMAX/SKID 
(http:\-//www-hpcc.astro.washington.edu/tools/SKID).
We confirm a good agreement between the cumulative cluster mass 
functions $n(>M)$ obtained using group masses from FoF ($\phi = 0.2$)
and HOP (suitably tuning its parameters), and that most massive HOP clusters 
show a systematic offset towards greater masses (\cite{Gov}).
The mass function found by DENMAX (again suitably tuned) finds a 
slightly smaller $n(>M)$, while the SO mass function mostly lays slightly below
all of them. Discrepancies are however small, in spite of being enhanced
by the reduced mass resolution in the test simulations. The results of
a detailed comparison among different cluster finders, applied to
higher resolution simulations, will be presented elsewhere, stressing
also morphological differences. 

The SO procedure, however, is the
one which provides clusters which appear to satisfy a precise virialization
requirement (a given overdensity in a sphere).
In this work SO was started setting $\phi = 0.2$, and used $N_f = 25$
(when using CDM particles only) or 75 (when also HDM particles are taken).

In fig.~\ref{fig5} we show the location and the mass of clusters in the same
simulation slices shown in fig.~\ref{fig1}. Circles are centered on cluster
centers of mass and their radii are proportional to cluster masses
obtained summing all particle masses within $R_s$ 
(the radius scale is set on graphic criteria and circles mostly exceed 
the projected physical volumes of clusters).
Two features can be easily seen even by eye: (i) The location of clusters
is similar in the three simulations, as is to be expected, as they are
started with the same random numbers. (ii) In average, MDM1 clusters are more
massive than TCDM, but not so numerous. Qualitative differences between
MDM2 and TCDM are not so striking.

\placefigure{fig5}

In fig.~\ref{fig6} we report the cumulative cluster mass functions, for TCDM 
and for all particles of MDM1 and MDM2 (open circles), 
compared with PS mass function curves, as given by eq.~(2.6), 
taking five equally spaced $\delta_c $ values between 1.4 
and 1.8 ($\delta_c = 1.686$ corresponds to an isolated spherical protocluster 
collapse). These plots show a substantial agreement
between theoretical expectations and simulation outputs,
confirming previous results by \cite{walter}. Some discrepancies,
that will be discussed soon, can be fairly easily understood and
are really minor effects, when one considers how complex is the
problem that the theoretical PS expression tries to face.

Discrepancies can be expressed through the dependence of $\delta_c$
on scales and models. For all models $\delta_c$ decreases when passing
from smaller to greater scales. However, while for MDM1 it starts
from values exceeding 1.8, at scales $\sim 10^{14} h^{-1} M_\odot$,
for MDM2, 1.8 is never exceeded and, for TCDM, the start occurs
for $\delta_c \sim 1.7$. From $ 10^{14} h^{-1} M_\odot$ to $\sim 10^{15}
 h^{-1} M_\odot$, $\delta_c$ approximately decreases by 0.1 (some
irregularity seems however present in MDM2). At even greater scales
a further excess is met, leading to values $\delta_c$ up to $\sim 1.45$, for 
TCDM. Although the final $\delta_c$ is smaller for MDM1 and MDM2, the 
excess of large mass clusters met in MDM1 is greater. Similar features 
were already 
outlined by previous analyses of pure CDM simulations (see, e.g., \cite{Ma1}, 
for low masses, \cite{Gov}, for the large mass end). Our analysis
shows that they are present also in
mixed simulations; furthermore, we find a decrease of cluster numbers,
compared to PS expectations, for mixed models. 
The scale dependence is thought to arise because of deviations from spherical 
isolated growth. The smaller
number of clusters we find in mixed simulations, instead, can be
attributed to a slower gravitational growth, caused by the enhanced
thermal velocities in one of the matter components.

\placefigure{fig6}

However, when mixed model predictions are to be compared with optical
observations, the above analysis is unsuitable, as already argued in
the Introduction. With the resolution allowed by our simulations,
hydrodynamical effects are expected to be negligible. Baryons emitting
light are therefore to be sought where they are initially set, when
CDM {\sl particles} were distributed at $z_{in}$, using the (already 
coincident) transfer
functions of CDM and baryons and granting them an overall density parameter
$\Omega_c + \Omega_b$. Only hydrodynamics could separate CDM from
baryons. Let us remind that, instead,
HDM particles were initially distributed according to
a (still) different transfer function, their evolution permanently
resented initial thermal velocities, which baryons certainly do not have,
and they have a $z=0$ spatial distribution different from CDM particles,
namely in those sites where non--linearity forced CDM to evolve so
rapidly that HDM could not overcome its pace and reach it.
In the Introduction we reported on attempts by previous authors
to locate baryons in non--hydro mixed simulations. Suitable prescriptions
were suggested to disentangle gas from CDM, in simulations with
higher resolution than ours; however, even with such greater
resolution, no doubt was cast on the fact that galaxies are CDM tracers.

At $z = 0$, discrepancies among transfer functions,
set by linear evolution, are marginal. At the linear level, therefore,
the distributions of baryons and dark matter components are
already alike, down to galactic scales. Non linear evolution,
instead, is faster and the final distribution of
cold and hot particles, involved in non--linear processes, keep
different. In fig.~\ref{fig7}, \ref{fig8} and \ref{fig9},
we show several model clusters, at different mass scales and with 
different morphologies, for TCDM, MDM1 and MDM2, respectively.
In fig.~ \ref{fig8} and \ref{fig9}, where
2--dimensional projections of 4 clusters from MDM1 and MDM2 
simulations are shown, such differences are evident. In fact,
an eye inspection of CDM and HDM distributions shows
that dense knots are mostly populated by CDM particles, while the
HDM particle distribution is significantly smoother. E.g., the second cluster
in fig.~\ref{fig8} has a triple structure in CDM, which
is almost absent in HDM. A similar situation holds for the third one,
whose double structure is erased at the HDM particle level. The last two
clusters are lighter ones, but it is still visible how central knots
are essentially made by CDM particles. Similar features are visible
(and perhaps stronger, probably because HDM particles are even lighter) 
for MDM2. Multiple features and knots, present in CDM distributions,
are attenuated or vanish in HDM. 
Similar properties can be observed in Broedbeck et al. (1998) visualization
work, based on MDM PM simulations performed in a smaller volume (a 50$\, 
h^{-1}$Mpc side cube), but with a force resolution similar to
ours and a better mass resolution. Accordingly, their cluster
sample is 40 times less numerous than ours.

It should be mentioned, however, that multiple clusters, such as those shown
in the second and third panels of fig.~\ref{fig8}, are, in a sense, 
survivors. In most cases, CDM bridges are not adequate to unify the smaller
condensations. Then we identify several smaller clusters instead
of a big single one. But, more often, some of the smaller condensations pass
below the lower mass threshold and do not contribute to the
cluster mass function. These effects are unique to identifying clusters
with CDM particles only in mixed simulations; if we use all particles,
we find a consistent extra number of multiple clusters like
the ones shown in fig.~\ref{fig8}.

\placefigure{fig7}
\placefigure{fig8}
\placefigure{fig9}

We conclude that most features expected in mixed model simulations are
indeed present. Let us now quantify them.

\section{Cluster mass estimates}

In order to pass from cluster masses as sums of particle
masses, to cluster masses comparable with optical data, two operations
are to be performed. We first apply the SO algorithm
to cold particles only. Once cold particles within $R_s$ are selected 
(let $M$ be their total mass), we evaluate their rms velocity $\sigma_v$
and use the relation
$$
M_V = {2 \over 3} { \sigma^2_v R_s \over G }~,
\eqno(6.1)
$$
to work out the virial mass $M_V$. Here
$G$ is the gravitational constant; the factor 2/3 implies that
particle distribution is assumed to be approximately isothermal
(Peebles 1993, \cite{Cen}).

In fig.~\ref{fig10} we compare the behaviours of $n(>M)$ obtained
for MDM1 and MDM2 in three different ways: (i) using all particles to define
clusters and summing their masses; (ii) using only CDM particles to
define clusters and summing their masses; (iii) using CDM particles
to define clusters and eq.~(6.1) to work out $M_V$, as cluster mass, from 
CDM particle velocities. For the sake of comparison, in fig.~\ref{fig11},
mass functions, obtained according to the (ii) and (iii) criteria,
are shown also for TCDM. The $n(>M)$ behaviour, built to give the
best approach to cluster mass estimate from data, is the (iii) one.
In general, passing from (i) to (iii), the
decrease of the cumulative mass function, at the low mass end, approaches
half order of magnitude. The difference becomes gradually smaller
for greater and greater mass scales. The behaviour, however, depends
on the matter mix and it may be premature to give an expression for cluster 
abundance reduction, on various mass scales. Moreover, the method 
used to select particles belonging to a cluster, which has already
an impact in defining masses according to the (ii) criterion, bears
an even greater effect on the (iii) criterion: here, adding or subtracting
a few high--speed particles may significantly change $\sigma_v$; the 
operational assumption of spherical symmetry in the SO and similar 
procedures, in principle, might cause significant $M_V$ shifts, for
intrinsically non--spherical clusters. (Let us mention that procedures
allowing non--spherical clusters may have even more serious difficulties 
to select really bound particles). However, approximately, on 
mass scales $\sim 4$--$5\, h^{-1} 10^{14} M_\odot$, the reduction
factor from (i) to (iii) is 0.6--0.8. 

\placefigure{fig10}
\placefigure{fig11}

Cen (1997) applied eq.~(6.1) to CDM simulations, but then
went further, mimicking a number of observational biases and, in particular,
evaluated $\sigma_v$ from all particles in
observational {\it cylinders} of radii $R_i = 1$ or 2$\, h^{-1}$Mpc and
constant depth $D = 100\, h^{-1}$Mpc, centered on cluster images.

Here we compare $M_V$ with $M$, directly estimated in spheres of radius $R_s$
(instead of cylinders), aiming to disentangle observational biases
on $\sigma_v$, arising from projection effects due to particles in cylinders
(interlopers),
from possible real physical features. In principle, for pure CDM, $M_V$ 
and $M$ should coincide. On the contrary, Cen (1997) found $M_V$ values 
slightly exceeding $M$, at small mass scales. The excess disappeared
at the top mass end, where he found $M_V$ values smaller than $M$.

This trend certainly includes projection effects. In fact, the expected number
of interlopers is proportional to the volume $\pi R_i^2 D$, 
and is independent from the mass $M$ of the cluster. On the contrary, 
within a fixed $R_i$, just for geometrical effects,
the number of galaxies belonging to the cluster is 
expected to be $\propto M^{1/3}$. Observational biases for top mass clusters
should therefore be smaller. 

This point is apparently confirmed by fig.~\ref{fig12}, where the 
straight line, in all panels, yields $M = M_V$. The right panel shows
what happens for a pure CDM model. At lower masses, the number of clusters 
is greater, and the range of $M/M_V$ ratios is wider. In spite of that, 
it seems clear that an underestimate of masses is present all over the 
scales, if eq.~(6.1) is used. The best--fit straight
line through $M$,$M_V$ points has a slope $0.89 \pm 0.013$.
This value, safely below unity, confirms that the effect found by
Cen (1997) is not only due to projections. Let us outline that the
best--fit line can be hardly found by eye, from fig.~\ref{fig12}, where the 
distribution of points on vertical bars at fixed $M$ cannot be spotted.

This effect has also an impact on the cluster mass function for TCDM.
In fig.~\ref{fig11} we show $n(>M)$ for TCDM, obtained using either $M$
or $M_V$ as cluster masses. The latter curve is smaller by a factor
$\sim 0.85$. In our opinion, the latter value are a better approach
to observational data.

\placefigure{fig12}

Let us now compare the effect found for pure CDM with what happens in 
mixed models. In spite of the noise due to peculiar cluster
features, the left panel of fig.~\ref{fig12} (MDM1) shows that the data
trend is steeper than for CDM. At the low mass
end, virial masses seem consistent with particle masses. In average,
this no longer seems true in the top mass range, even though the noise is 
large. A numerical fit gives a slope $1.21 \pm 0.020$.

The central panel refers to MDM2 and its features are somehow
intermediate. At low $M$, $M_V$ values
have a deficit which looks much alike the one in pure CDM. This
deficit decreases for greater $M$ and, at the top mass
end, there seems to be a fair coincidence between $M$ and $M_V$.
The best--fit slope is $1.05 \pm 0.016$. This is intermediate
between CDM and MDM1, but statistically distinct from both.

The first and main conclusion from the above arguments is that
virial mass estimates give values whose relation with 
masses obtained summing all the particles
depends on the substance by which the real world is made.
The trend shown in fig~\ref{fig12} is confirmed if we focus
at the top mass end (see fig.~\ref{fig13}, which is analogous to 
a figure given by \cite{Cen}). By comparing the trend of $M/M_V$ in 
MDM with CDM models, there seems to be some evidence that, in average, 
the amount of HDM trapped inside a cluster of mass $M$ increases 
with $M$ itself. More HDM means a deeper potential well and also
more CDM can be trapped. Within this interpretation the difference 
between MDM1 and MDM2 is to be ascribed to the different amounts of HDM.

We can conclude that: (i) The cumulative cluster abundance,
that we expect to measure on optical data,
in the presence of a hot component,
is significantly smaller than standard PS estimates. The reduction amounts
to a factor $\sim 0.3$--0.4 at the low mass end and reduces to 
$\sim 0.6$--0.8, depending on the mix, on 
mass scales $\sim 4$--$5\, h^{-1} 10^{14} M_\odot$, where
cosmological models are usually compared with cluster observations.
(ii) We should not expect that this effect is covered
by projection effects. The discrepancy between $M$ and $M_V$, found
in Cen (1997) is confirmed also excluding them. Notice
also that the very cluster mass function for CDM has a significant
shift, when using virial masses. (iii) A contribute to the 
discrepancy between $M$ and $M_V$, which is found also in the CDM case,
is likely to arise from the assumptions that the matter 
distribution, inside clusters, is approximately isothermal. 
We plan to deepen this point
in a forthcoming analysis, studying cluster profiles and other
features of matter distributions in clusters.

In fig.~\ref{fig10} we compare simulation mass functions with data.
Open circles are Biviano et al. (1993) data points,
filled circles are Girardi et al. (1998) data points. The latter analysis
uses a wider sample and is corrected for
several biases. If the same corrections are performed
on the older data set, however, there is
an excellent fit between the two observational outputs.
The biases considered were:
(i) Subclustering, which might vary individual cluster masses
by a large factor, but is found not to have significant effects on the
overall function, which is quite alike both if a $\sim 10\, \% $ of
substructured clusters is included or excluded. (ii) Interlopers, whose
exclusion causes
an average reduction of cluster masses by a factor accounting
for half of the discrepancy. (iii) Boundary effects, which
account for the rest of the shift in cluster masses.

Interlopers are clearly absent in simulation analysis.
However, the assumption of isothermal particle distribution in clusters is 
admittedly just a basic approximation. The analysis performed on data, taking
into account individual galaxy points as well as boundary effects, 
should be fully repeated on
mock catalogs built from simulations, to improve the comparison level.
This might even cause mass corrections similar to
those due to boundary effects, which cover half of
the shift between the two sets of data points. 
Accordingly, in spite of the fact that the two data sets meet 
MDM1 and MDM2 mass functions, respectively, 
we should refrain from stating that this shows that MDM2 is a better data fit
than MDM1. TCDM, instead, is farther from data (see fig.\ref{fig11}).
Let us also draw the attention on the
reduction of $n(>M)$ at the low mass end, which is
mostly due to using CDM particles only, to define clusters.

\placefigure{fig13}

Our simulations will allow an analysis of clustering evolution with $z$,
in mixed models. In fig.~\ref{fig14} we show a preliminary but significant 
result, by comparing the cluster mass functions at $z=0$ and $z=0.8$.
As expected, the degree of evolution of the three models is similar and
an excess of clusters at the top mass end is however present. 
As can be seen from  fig.~\ref{fig14}, in MDM1 there are
2 clusters with mass $\sim 1.8 \cdot 10^{15} M_\odot$.
In MDM2 the 3 largest clusters have a nearly identical mass $\sim 9 \cdot 
10^{14} M_\odot$. In TCDM, instead, there is 1 cluster of $\sim 6.5
\cdot 10^{14} M_\odot$; all other clusters are less massive.
This behaviour is coherent with the spectral features of the
three models, but is also linked to the higher $\sigma_8$ value required
in mixed models, to fit the observed cluster abundance at $z=0$, over
scales $\sim 4$--$5\, h^{-1} 10^{14} M_\odot$. 

If we scale Donahue et al. (1998) findings to the simulation volume, we 
expect there $\sim 0.75$ clusters of
mass $> 1.4\times 10^{15} M_\odot$. This allows to suggest that, thanks to
the different spectral shape and to the required higher normalization,
mixed models, with $\Omega_o = 1$, may not be in conflict with
Donahue et al. (1998) data analysis.

\placefigure{fig14}

\section{Discussion}

In this paper we have reported the basic outputs of two large N--body 
simulations, for mixed models and compared them to a tilted CDM
simulation. All models are consistent with linear constraints. 
The numerical code was built by implementing Couchman AP3M program 
for a parallel shared--memory machine. This allowed to attain a 
significant dynamical range with reasonable computational times 
($\sim 1$ week).

We first reconstructed linear and non linear spectra, 
from particle distributions. At scales exceeding $\sim 10\, h^{-1}$Mpc,
non--linear effects are small and reconstructed spectra fit linear
expectations, even at $z = 0$. Spectra are plotted against reconstructed
APM 3--dimensional data; the best fit, at large scales, is given by
MDM1, which nicely passes through observational bars. 
It has been known since several years that mixed models
provide a good fit of several features of LSS (\cite{bono3,liddle96,
prima95,smith98,gross98,gawi98}). Recent work (\cite{Pie,Bono1}) has shown
that the fit of galaxy spectrum data, about
the $\sim 200\, h^{-1}$Mpc peak, improves if a 
stronger bending of the spectrum, arising from lighter HDM particles,
is partially compensated by a blue spectral index.
Both such ingredients rise the CMB angular spectrum $C_l$,
which can therefore fit both COBE and more recent experiments
(\cite{Whi,Net}). E.g., fig.~3 shows that MDM2, which has a smaller $\Omega_h$
and $n$ just above 1, is not such a good fit to APM data as MDM1. Its
score, however, is better than TCDM. (Moreover, MDM2 is still
a good fit to $C_l$ at large $l$ values, while here TCDM fails.)
As far as the
non--linear behaviour is concerned, we found excellent fits with 
Peacock $\&$ Dodds (1996) expression. Mixed models, however, give non linear 
features slightly, but systematically, exceeding their expression.

Our simulations allowed us to follow the evolution of CDM and HDM
components, while clusters form and in the field between
different clusters. One might expect that the fraction of $\nu$'s,
captured in a cluster, increases with the deepening of its potential well.
This is neither what we see in the simulations, nor was it seen
in previous works (see, e.g., Klypin et al. 1997, and references
therein; let us also draw the reader's attention on the visualization
work by Broedbeck et al. 1998). Rather, $\nu$'s tend to
remain on the outskirts of clusters, whose knots, in average,
are partially $\nu$--emptied. The point is that gravitational growth,
in highly non--linear sites, can be so fast that HDM fluctuations
can hardly cover the gap they still had at $z_{in}$, in respect to CDM,
and that they almost cover in the linear growth.

The main concern of this work, however, was the cluster mass distribution.
The expectations outlined in the Introduction section were confirmed
by our numerical analysis. In particular we found that the actual
number of clusters {\sl produced} by a mixed model is significantly
smaller than the amount predicted by PS estimates. 
On a scale $\sim 4 \cdot h^{-1} 10^{14} M_\odot$, as the one usually
adopted to test models using their transfer function, the reduction
factor is $\sim 0.5$--0.8.
Simulations were actually compared with optical observational data
(Biviano et al. 1993, Girardi et al. 1998). The latter data are obtained
from a much wider sample, but most discrepancies arise just from
the treatment of observational biases. The mass function obtained from
simulations directly excludes some of such biases, e.g., the
presence of interlopers. Some other corrections might however be
needed before the comparison is safe, which might cause a further
decrease of estimated cluster masses. Accordingly, although 
MDM2 seems quite consistent with observations and provides an excellent
fit of data at various mass scales, also MDM1 cannot be safely excluded.
PS estimates are to be decreased by a suitable (smaller) factor, also for
CDM, in accordance with previous findings of Cen (1997).

In this work we did not pursue a comparison of simulations with 
X--ray data on cluster abundance (see, e.g. Eke et al. 1998).   
They are known to be in fair agreement with optical data for
$M \magcir 4$--$5 \, h^{-1} \cdot 10^{14} M_\odot$
(see, e.g., fig.~9a in \cite{mazu96}).
At lower masses various problems are to be solved, both concerning
the completeness of real cluster samples and the dynamics of gas
in simulated clusters. One of our findings is the significant reduction
of optical cluster masses, when external layers, connected by HDM bridges,
are no longer included in the mass computation. In the analysis of
real cluster samples, galaxies belonging to such layers would appear
as interlopers and an eye analysis of fig.~7 of \cite{mazu96}
seems to confirm that including (at least some of) the interlopers
might rise the cluster mass function, approaching PS estimates.
This point was not deepened in this analysis, as the natural approach
is through the use of mock catalogs, as we plan to do in a forthcoming work.

Cen (1997) analysis of the relation between virial mass estimates $M_V$ 
and real masses $M$ is widened here to mixed models.
We find that the $M_V/M$ ratios  depend on the cluster
mass scale. In general, the relation between cluster
virial mass estimates and real cluster masses,
is likely to depend on dark matter nature.
The hypothesis of isothermal distribution of particles, inside
clusters, is also found to yield masses (slightly) smaller than real, in
pure CDM simulations. In mixed simulations
the discrepancy between $M_V$ and cold particle masses is smaller
than in pure CDM. On the contrary, taking into account also
HDM particles, the gap is substantially greater and shows a
significant scale dependence, although the detailed
trend depends on the substance of the model.

Let us finally outline that, at variance with what tends to happen for 
galactic mass systems, the abundance of early clusters, in mixed models,
tends to exceed CDM. In our volume of 360$\, h^{-1}$Mpc,
2 clusters with mass $> 1.8 \times 10^{15} M_\odot$ (with $h = 0.5$)
were found in the MDM1 simulation. Also MDM2 provides a
cluster abundance compatible with Donahue 
et al. (1998) observational findings. Such numbers, however,
are too small, both on the observational and on the simulation sides,
to provide real confirms of models.

\acknowledgments
We thank the consortium CILEA for allowing us to run a
former trial simulation free of charge and a particular thank is
due to Giampaolo Bottoni of CILEA, for his expert technical assistance.
Elena d'Onghia, who gave us useful suggestions for the 
parallelization of the PP part of the code, is
also to be gratefully thanked. Hugh Couchman is to be
thanked for making public his AP3M serial code and for some
private communications. Anatoly Klypin is also to be thanked for
accurate discussions of several points of this work.
We would finally thank Stefano Borgani for early discussions
on the topics of this work.

\vfill\eject

\vfill\eject

\begin{figure}
\caption{CDM particles in a $10 h^{-1}$ Mpc thick slice in MDM1, MDM2 and TCDM 
simulations. Half of mass points are shown.}
\label{fig1}
\end{figure}

\begin{figure}
\caption{Spectrum evolution in the three models.
On the first 2 lines, the 3 plots refer to the two components (cold and hot) 
of MDM1,2 and to their overall spectrum. The last plot refers to TCDM.
Spectra are shown at $z_{in} = 10$, at $z = 3$, $z = 1.13 $ and 
at $z = 0$. No correction for the lack of small scale power is 
performed on these plots.}
\label{fig2}
\end{figure}

\begin{figure}
\caption{Spectra of the three models at $z=0$.
Solid curves give the linear power spectrum and the spectrum corrected 
for non--linearity, according to Peacock $\&$ Dodds (1996). Empty 
squares yield the
simulation spectra corrected for CIC (see text). Circles with
2$\, \sigma$ errorbars are the power spectrum measured from the APM survey.}
\label{fig3}
\end{figure}

\begin{figure}
\caption{Magnification of non linearity onset.
Symbols are as in fig. \protect\ref{fig3}. Only a part of the points yielding the
CIC corrected spectra are shown, to avoid graphic confusion. Notice that MDM 
simulation points systematically exceed the Peacock $\&$ Dodds (1996) curve,
although by a small amount. The shift between total (empty squares) and CDM
(filled squares) spectra is quite small.}
\label{fig4}
\end{figure}

\begin{figure}
\caption{On the same slices as in fig. \protect\ref{fig1} we map
the position and the masses of clusters. Circle centers are
cluster centers of mass. Circle radii are proportional to cluster masses.
}
\label{fig5}
\end{figure}

\begin{figure}
\caption{Cluster cumulative mass functions for the 3 models.
Empty spheres are obtained from simulations. Lines give the
expected PS mass functions for 5 equally spaced $\delta_c$ values, ranging
from 1.4 at the top to 1.8$\, $ at the bottom. 
Mass functions for mixed models are
worked out using both cold and hot particles.}
\label{fig6}
\end{figure}

\begin{figure}
\caption{We show the x--y and z--y projections 
of 2 clusters in the TCDM simulation.
Masses are shown aside to each plot.}
\label{fig7}
\end{figure}

\begin{figure}
\caption{We show a projection of 4 clusters in the MDM1 simulation.
Cluster masses are shown aside to each couple of plots.
Each couple gives CDM and HDM particles, separately. To facilitate
a comparison, only half of HDM particles are shown. Notice how
structures are smoother or even disappear in HDM.
Notice also that MDM clusters have more inner structure than TCDM ones.
}
\label{fig8}
\end{figure}

\begin{figure}
\caption{We show a projection of 4 clusters in the MDM2 simulation.
Masses are shown aside to each couple of plots. An eye inspection confirms
the features outlined for MDM1 clusters.}
\label{fig9}
\end{figure}

\begin{figure}
\caption{Cluster cumulative mass functions for MDM1 and MDM2 (indicated by 
number labels) if cluster masses are obtained using all particles (dashed 
line), only cold particles (dotted line), or virial masses (solid line).
Observational 1--$\sigma$ error bars from Biviano et al. (1993; open circles) 
and Girardi et al. (1998, filled circles) are also shown. Recall that
the latter data set is based on the wider ENACS sample and is
suitably cleaned from interlopers and other biases.
}
\label{fig10}
\end{figure}

\begin{figure}
\caption{Cluster cumulative mass functions for TCDM 
if cluster masses are obtained summing particle masses (dashed line) or using 
the virial mass $M_V$ (solid line).
Data points and bars are as in the previous figure.}
\label{fig11}
\end{figure}

\begin{figure}
\caption{Cluster masses estimated using virial theorem ($M_v$)
vs. masses obtained summing masses of the cold particles ($M$). 
Particles within $R_s$ are taken.}
\label{fig12}
\end{figure}

\begin{figure}
\caption{Cluster masses estimated using virial theorem ($M_v$)
vs. masses obtained summing particle masses ($M$). Points within
1$\, h^{-1}$Mpc from the centers of mass of the 30 most
massive clusters are considered.}
\label{fig13}
\end{figure}

\begin{figure}
\caption{
Comparison of the cluster mass function at $z=0$ and $z=0.8$ 
for the three models.  }
\label{fig14}
\end{figure}

\vfill\eject

\begin{table}
\begin{center}
\begin{tabular}{l|c|c|c}
\multicolumn{4}{c}{} \\
&\multicolumn{1}{c}{MDM1}&\multicolumn{1}{c}{MDM2}&\multicolumn{1}{c}{TCDM}
\\
\hline
 $\Omega_h$           &          0.26     &      0.14    &     ---- \\
 $m_\nu/$eV           &         3.022     &     1.627    &     ---- \\
 $\Omega_b \cdot 10^2$ &         6.8      &       9      &      6   \\
 $n$                  &          1.2      &     1.05     &     0.8  \\
 $Q_{PS,rms}/\mu$K    &         12.1      &      13      &    17.4  \\
 $\sigma_8$           &         0.75      &    0.62      &    0.61  \\
 $\Gamma$             &         0.18      &    0.23      &   0.32   \\
 $N_{cl}$ (PS; $\delta_c = 1.69$)  &        14.      &      5.2      &     5.7\\
 $N_{cl}$ (sim)       &          10.      &     4.7      &    6.0   \\
 $L_\alpha $      &          1.3      &     1.2      &    1.3   \\
\multicolumn{4}{c}{} \\
\multicolumn{4}{c}{} \\
\end{tabular}
\caption{
Parameters of the models. All parameters listed are
either input parameters or quantities worked out from the
linear theory. The only exception is $N_{cl}$ (sim).
Mixed models were chosen in order to explore upper and lower
limits of possible cluster mass functions.
The normalization to COBE quadrupole was deliberately kept at the
$\sim 3 \, \sigma$ lower limit, in order leave some room to the
contribution of tensor modes, but keeping however consistent with data.
The expected interval for $N_{cl} $ is 4--6, but models with
$N_{cl}$ up to 8--10 cannot be safely rejected. The $2 \, \sigma$
lower limit for $L_\alpha$ is $\simeq 1.3\, $.
}
\label{tab1}
\end{center}
\end{table}

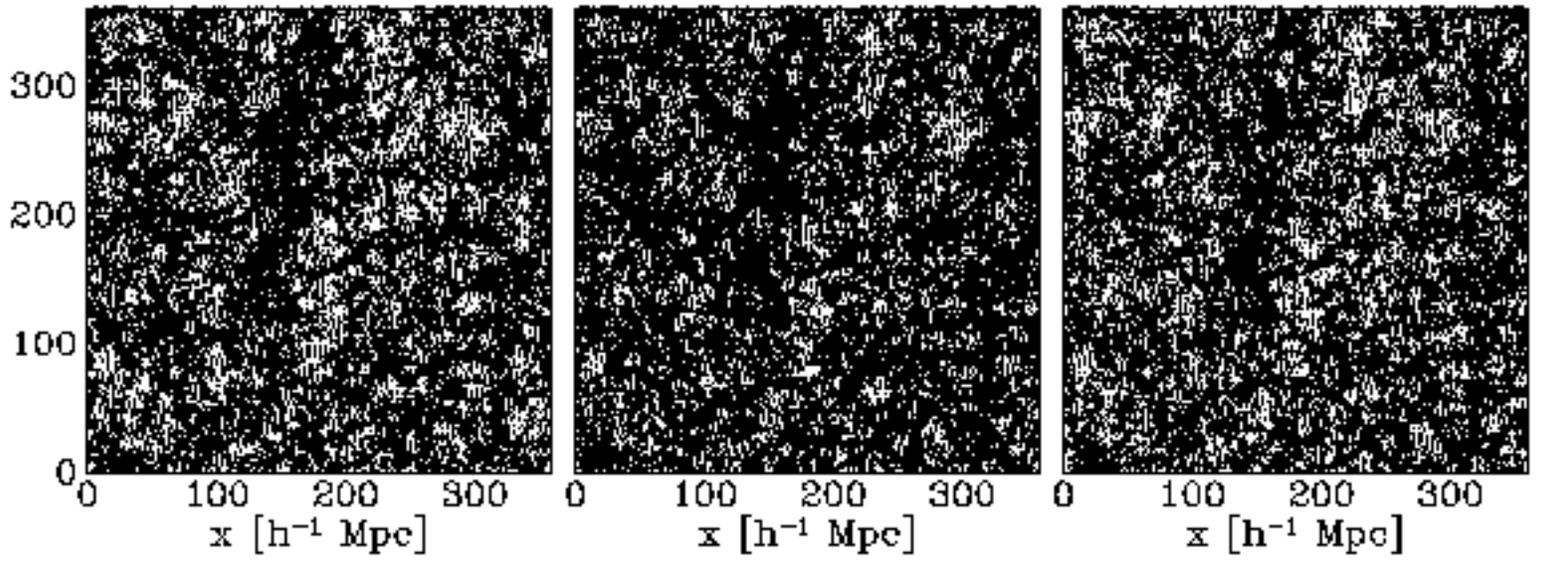
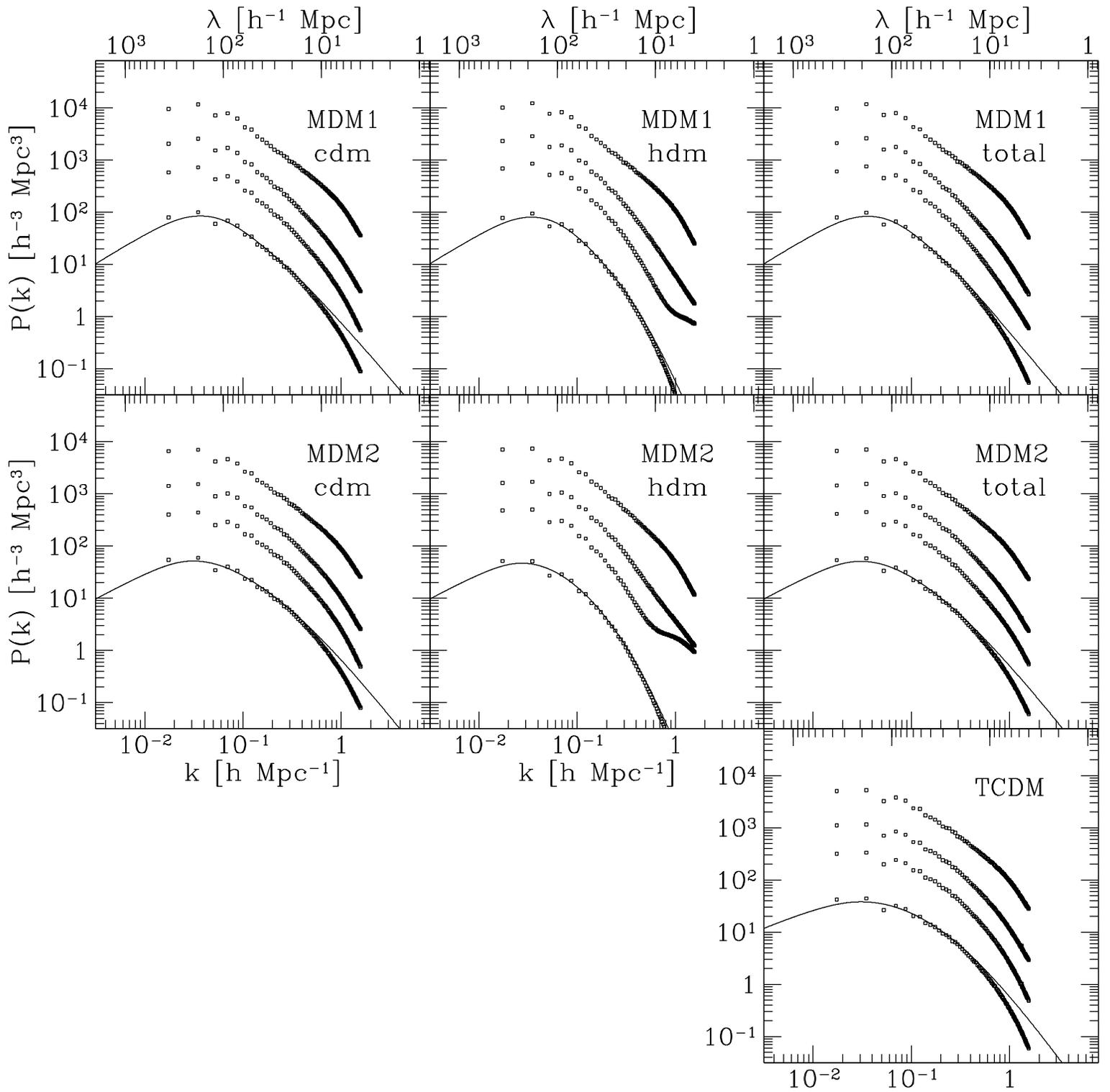
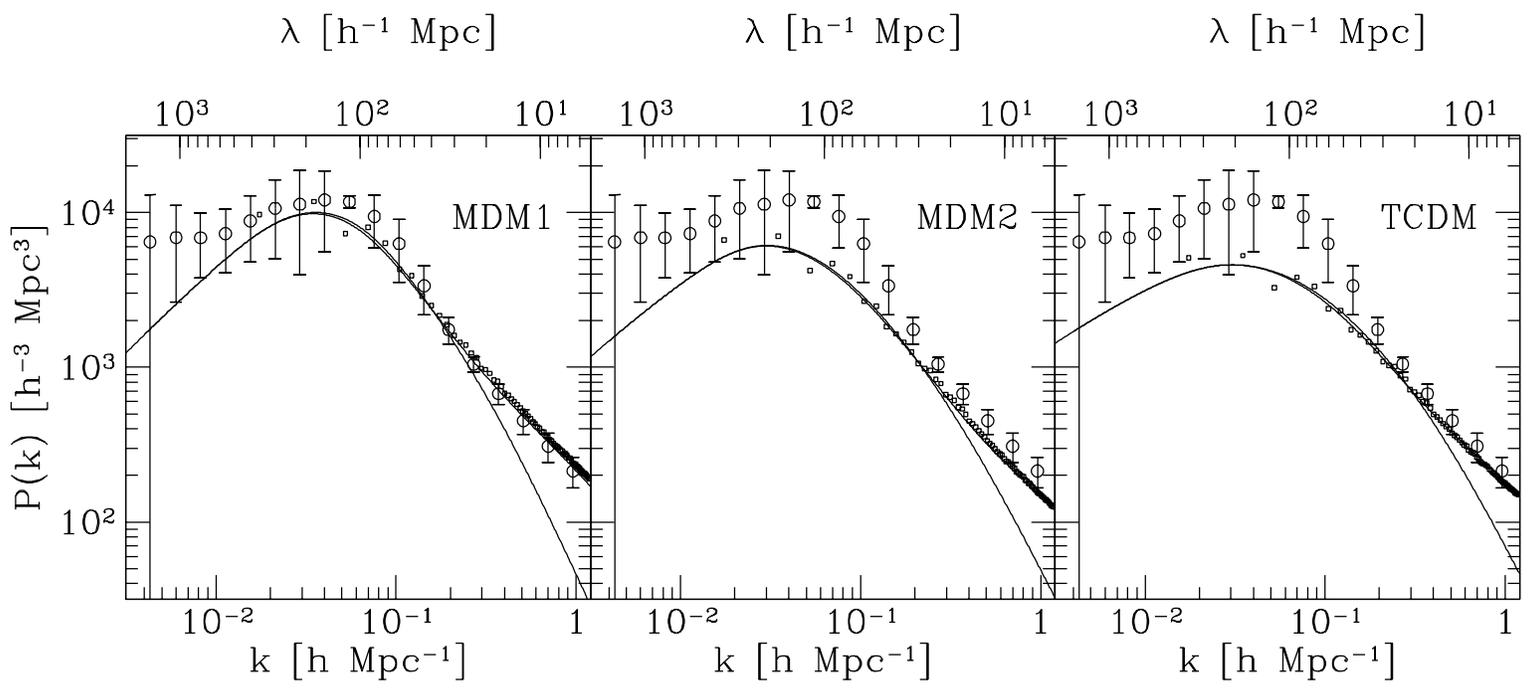
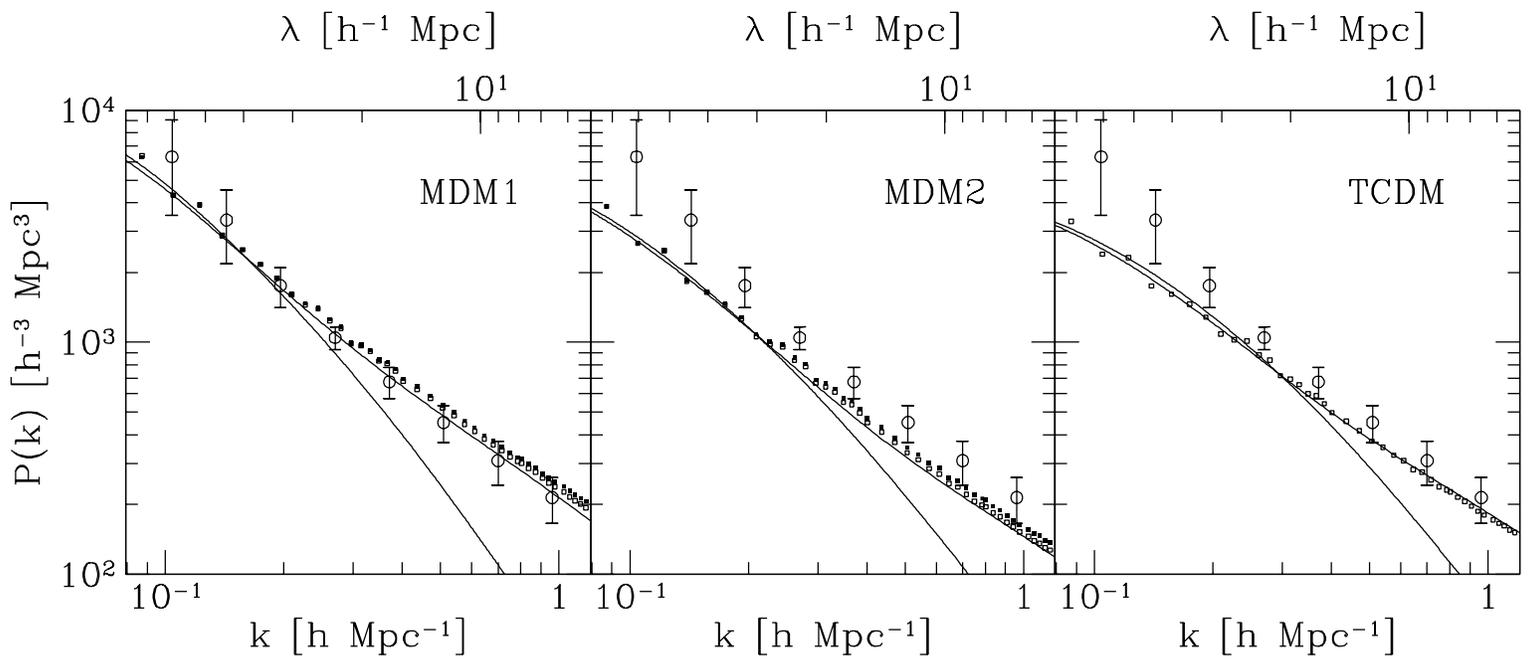
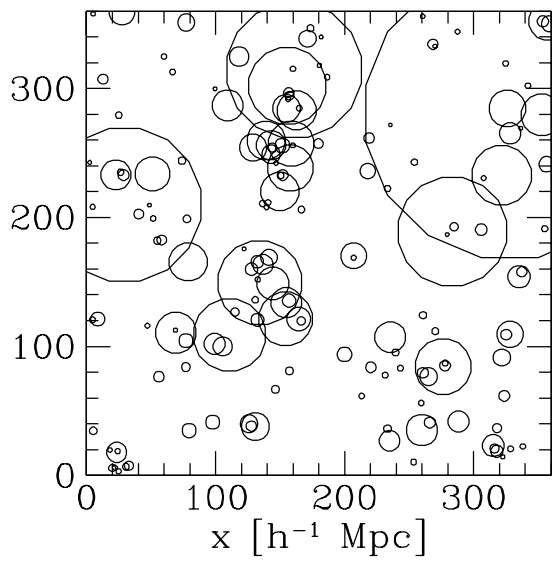
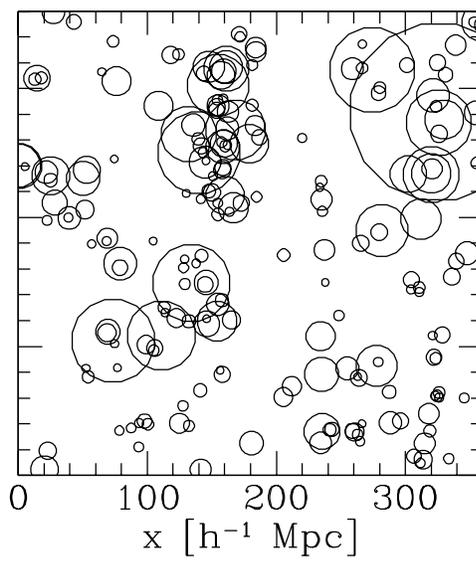
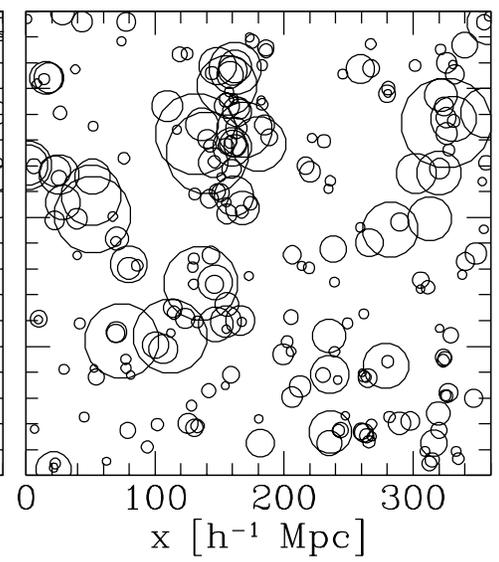
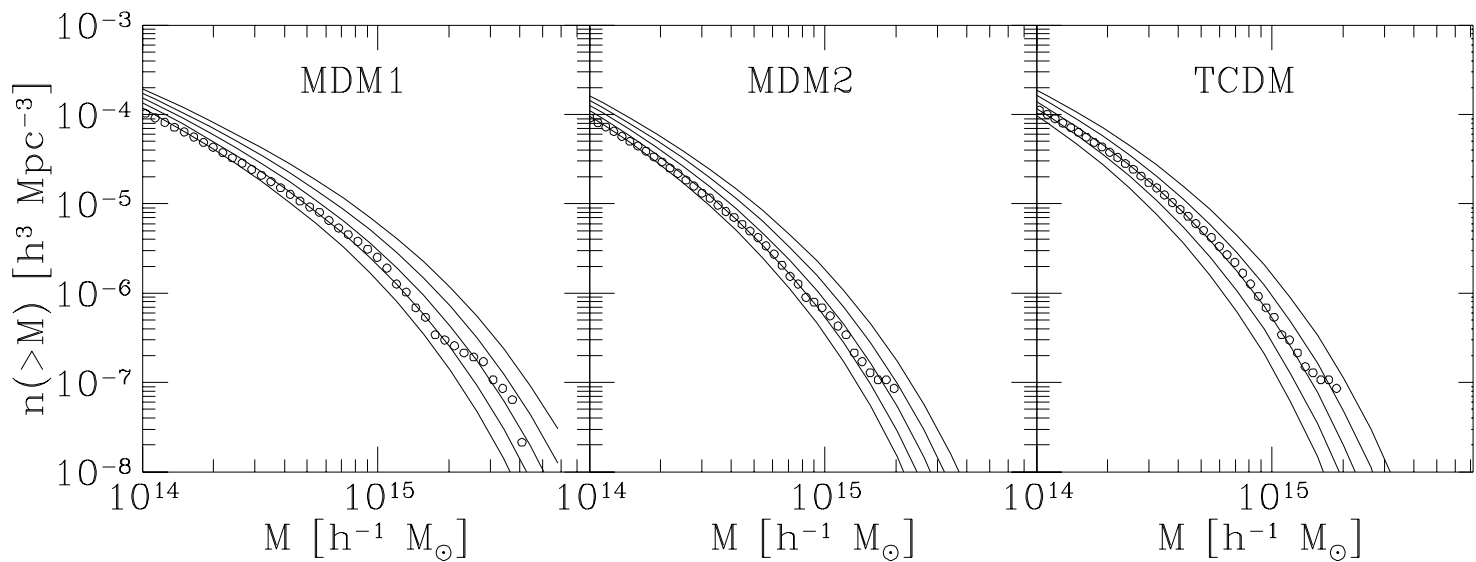
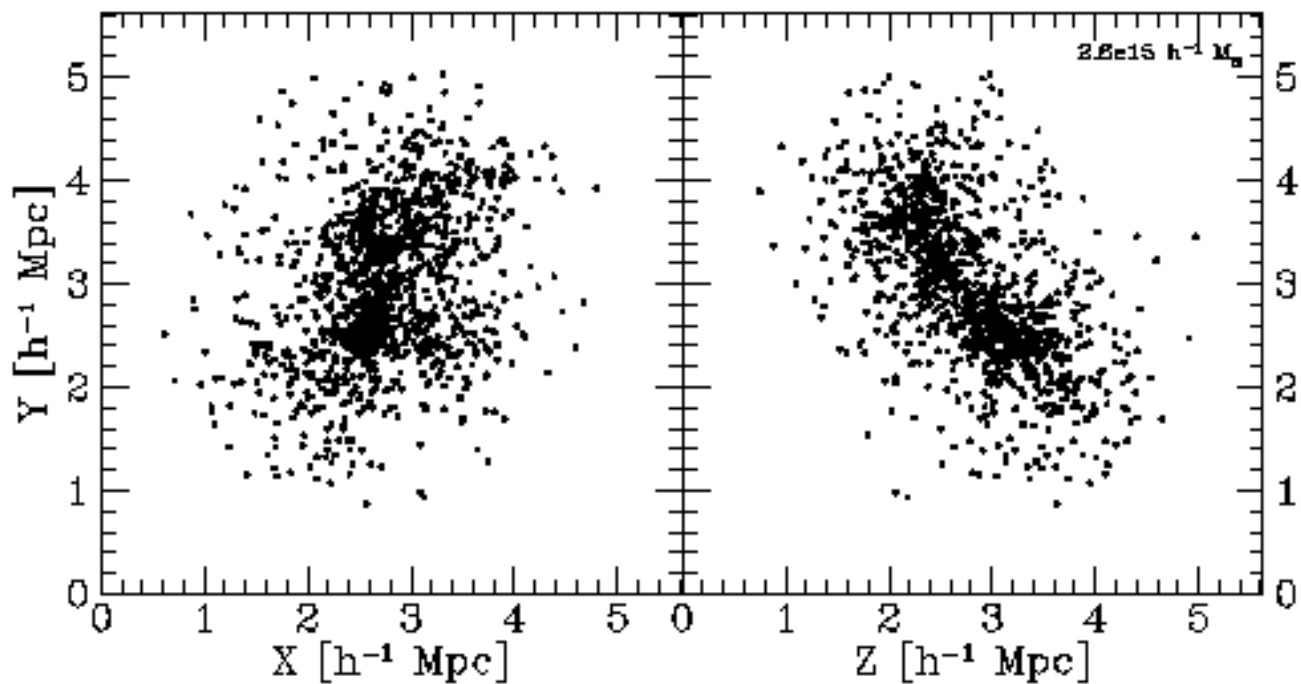
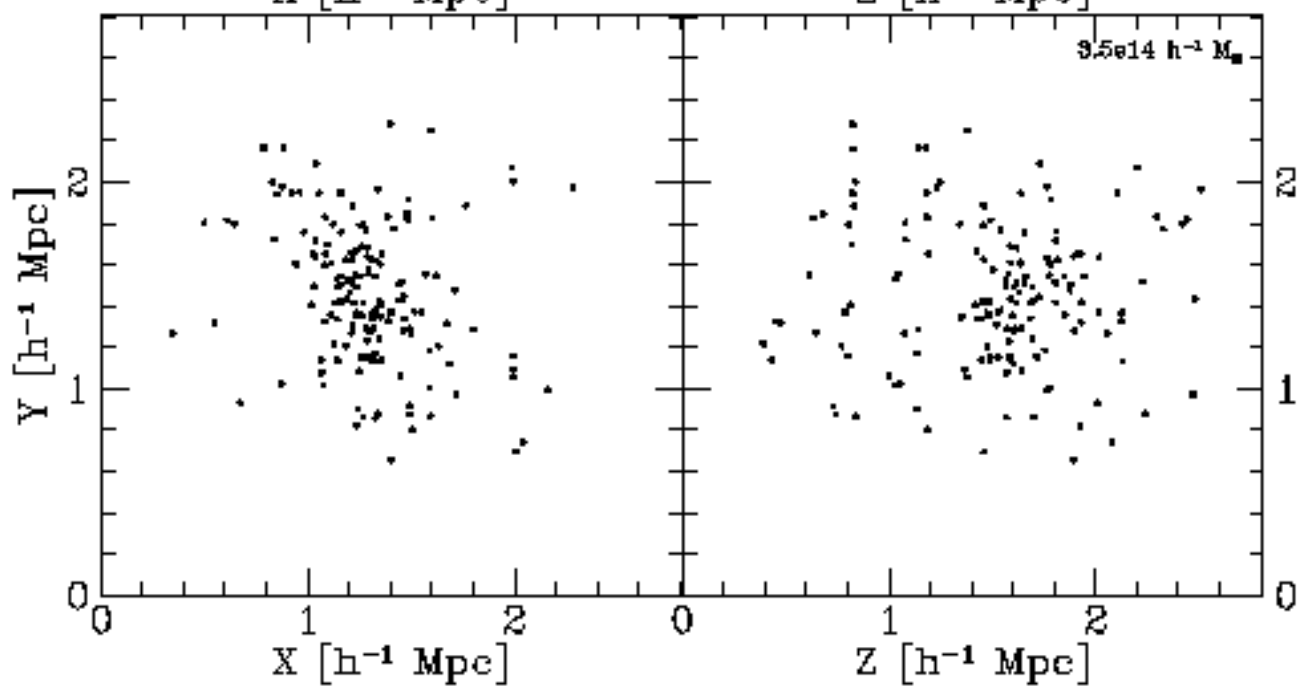
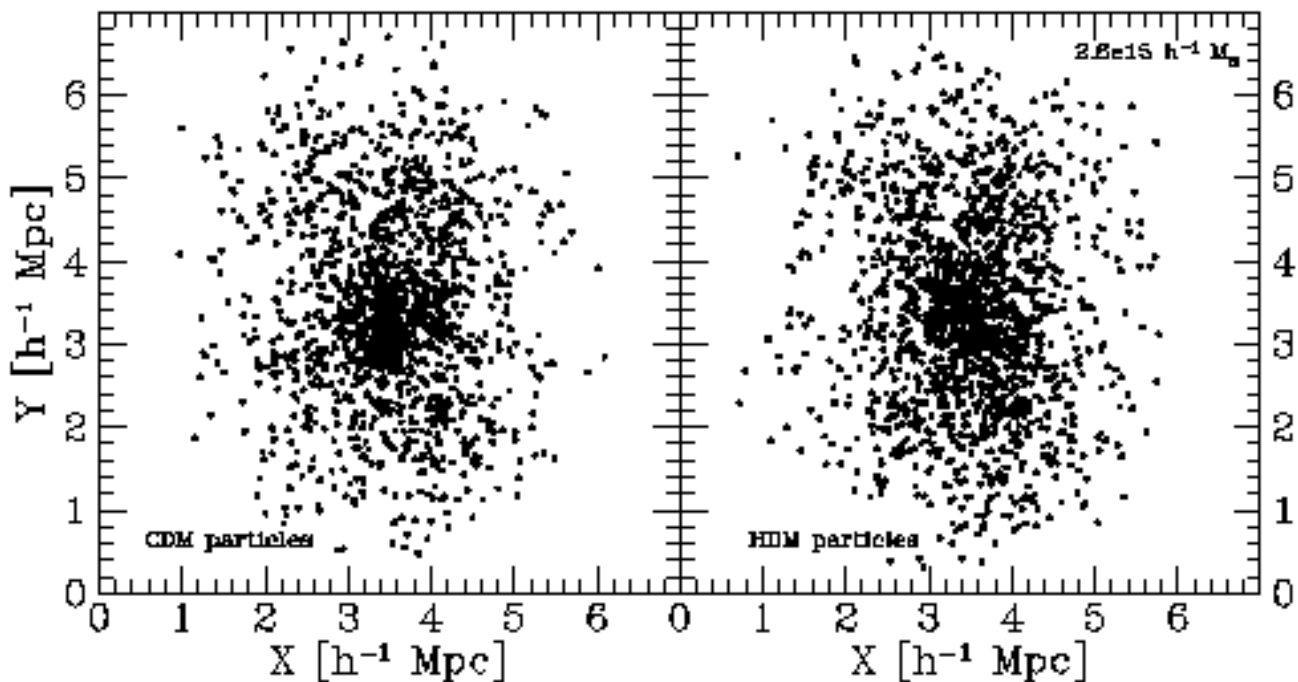
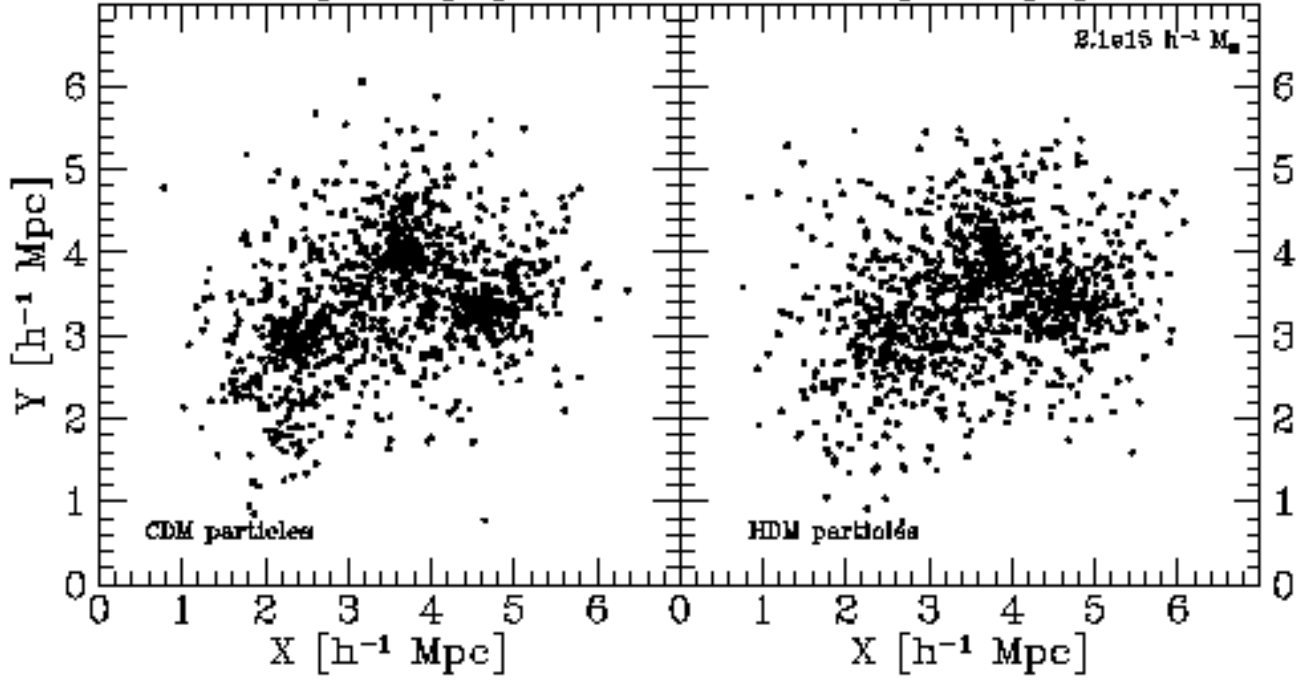
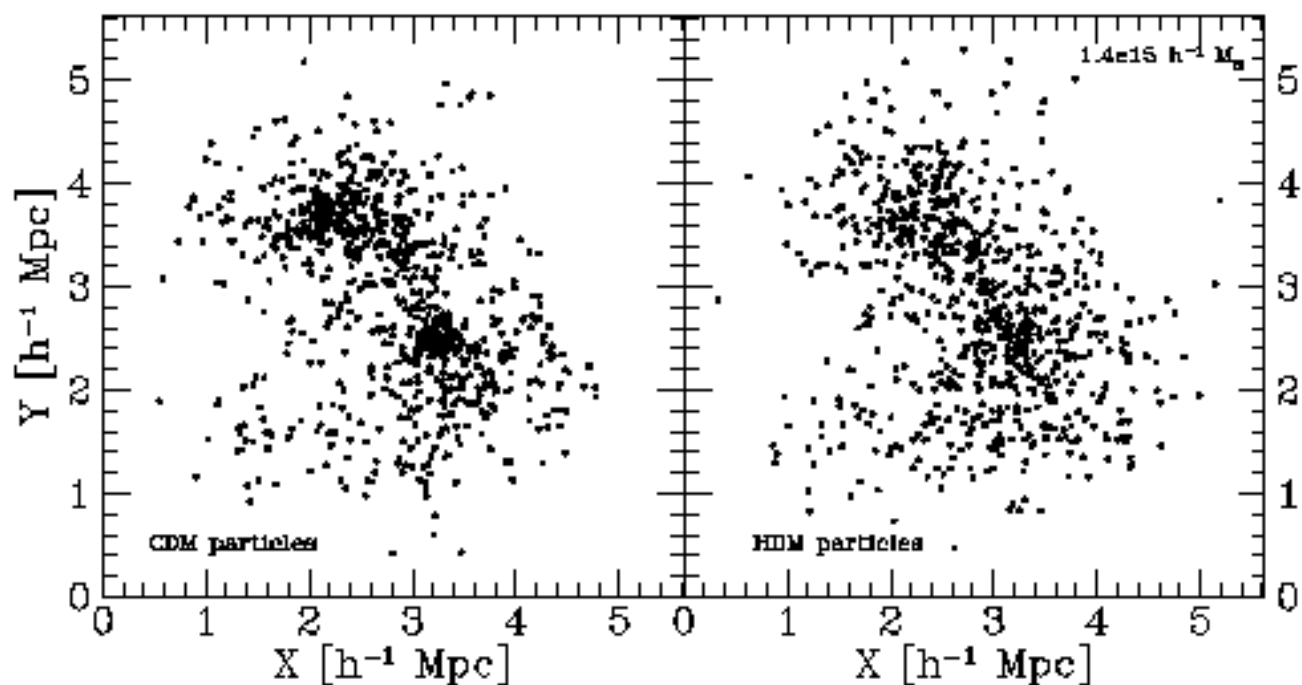
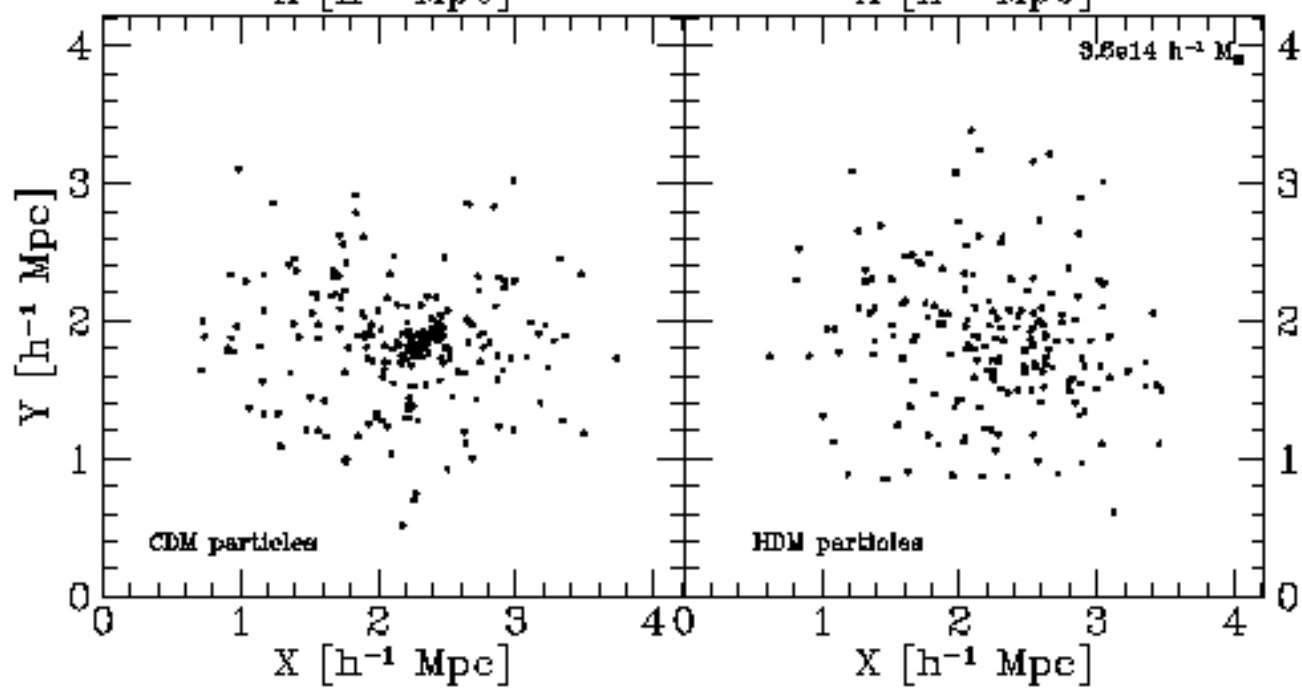
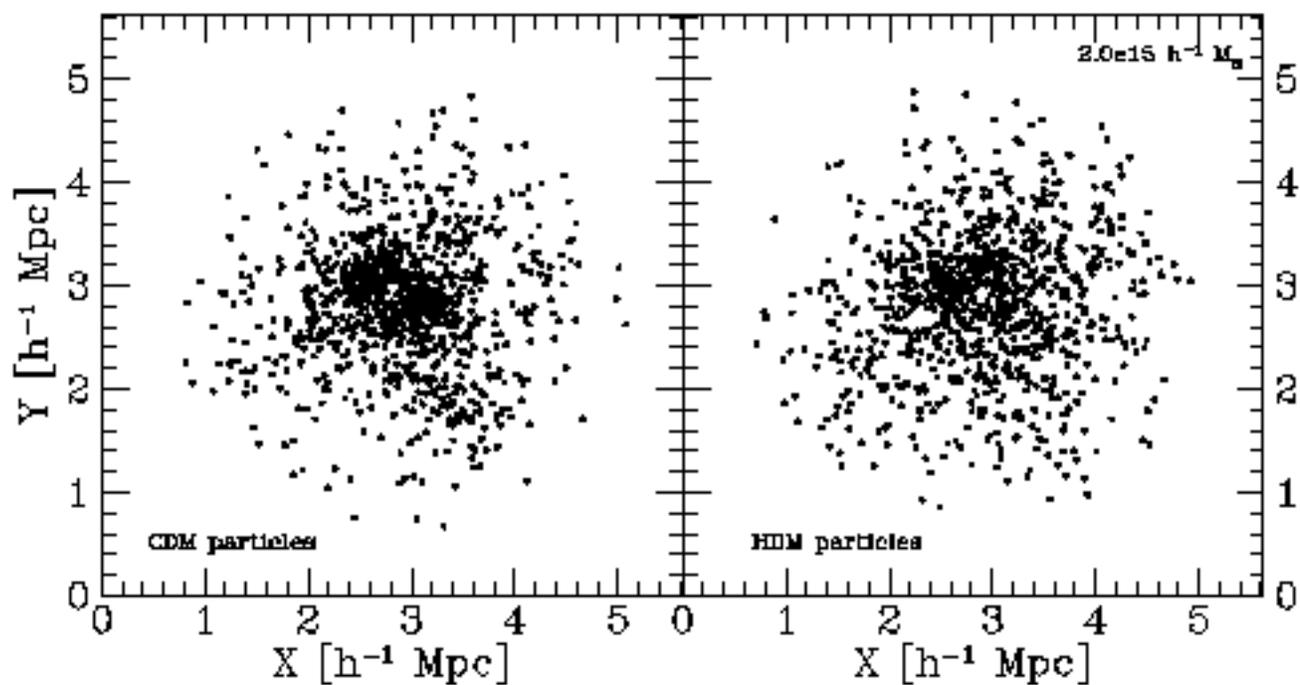
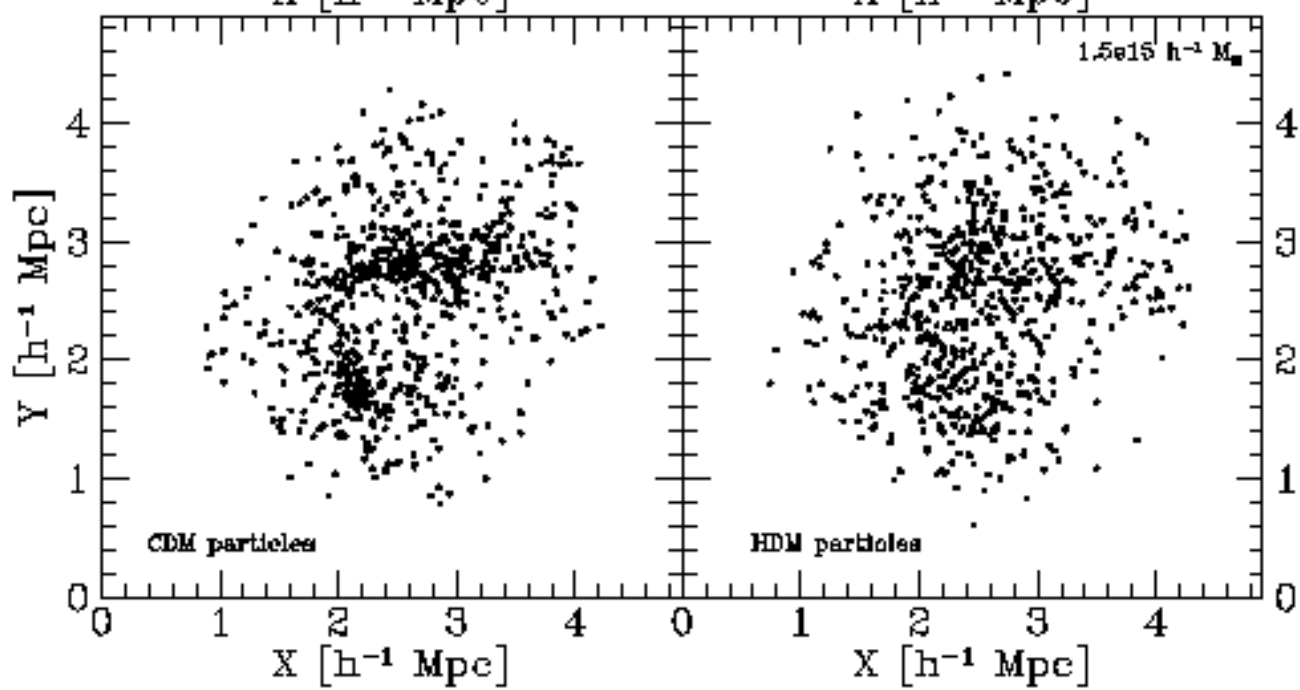
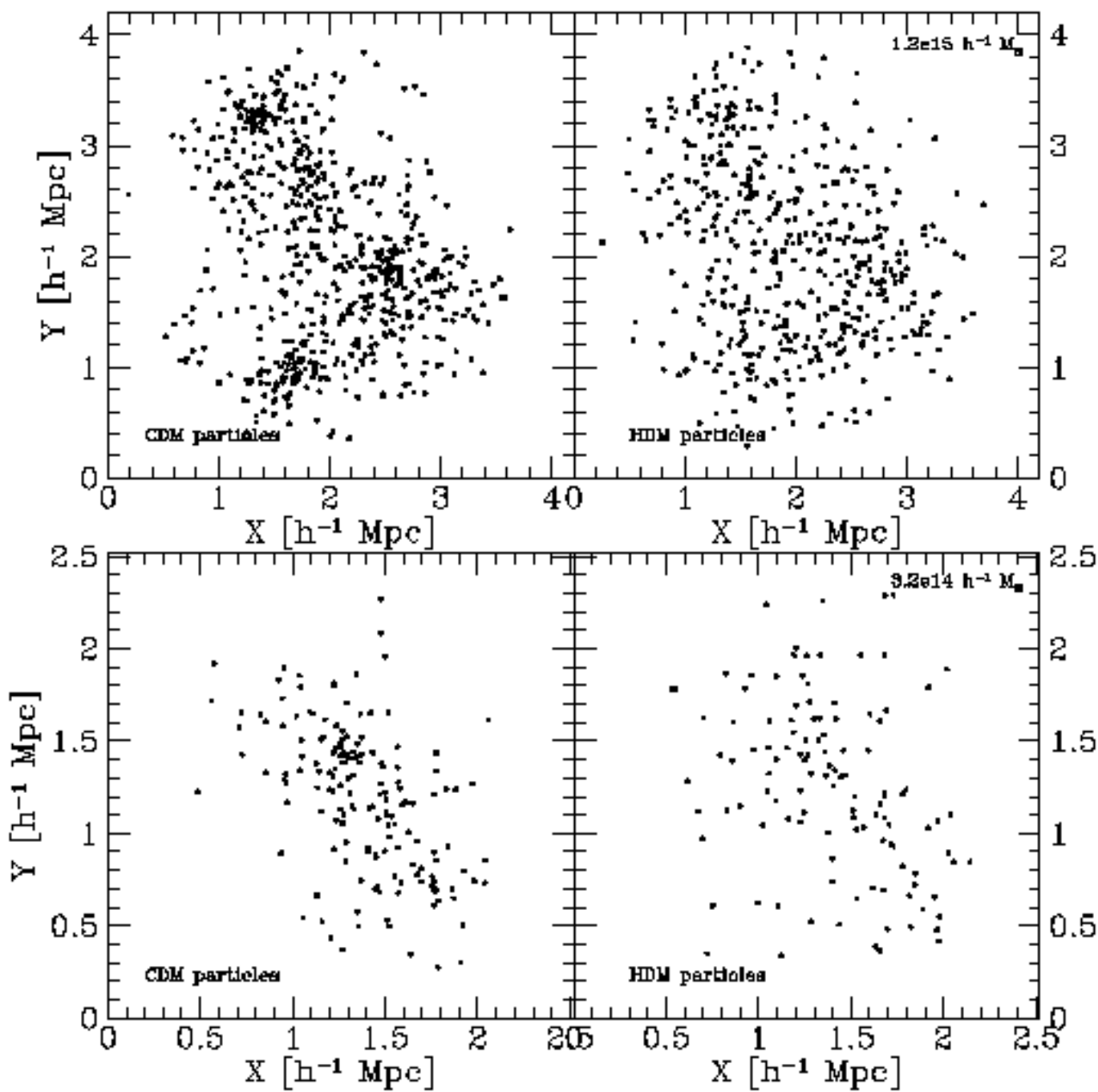
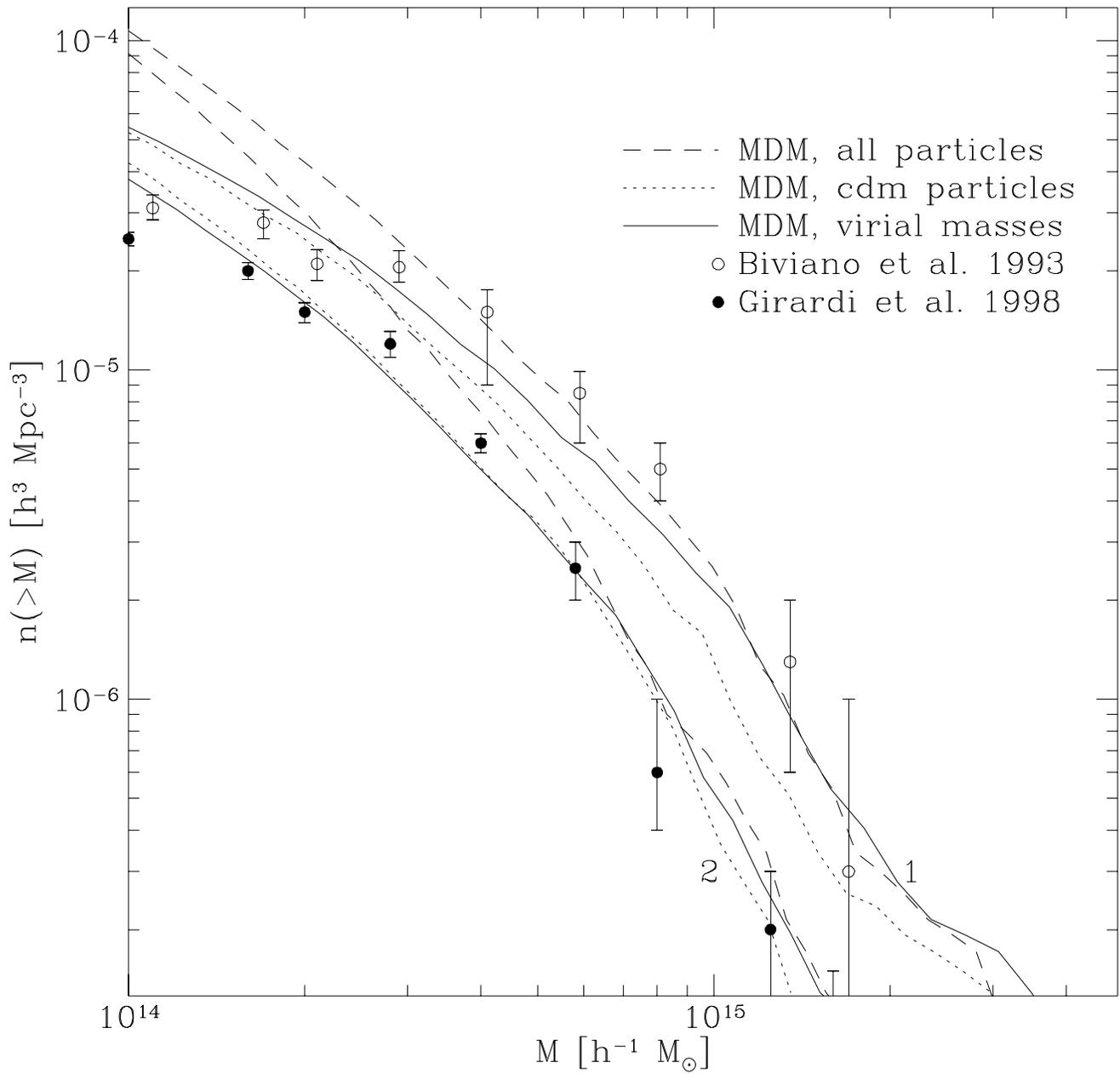
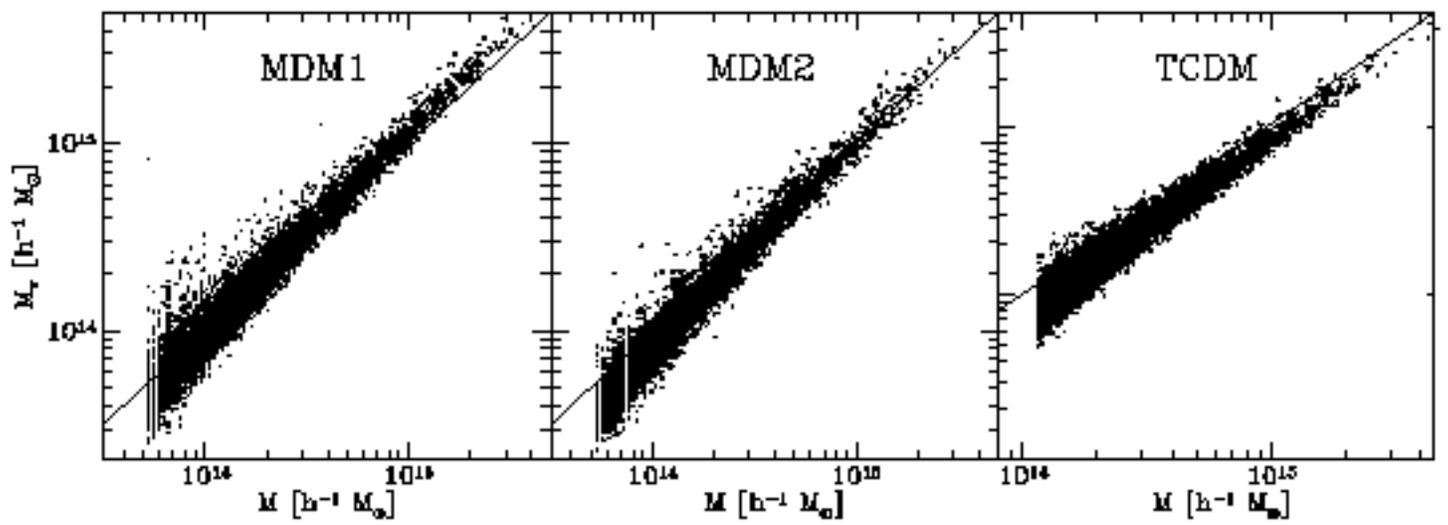
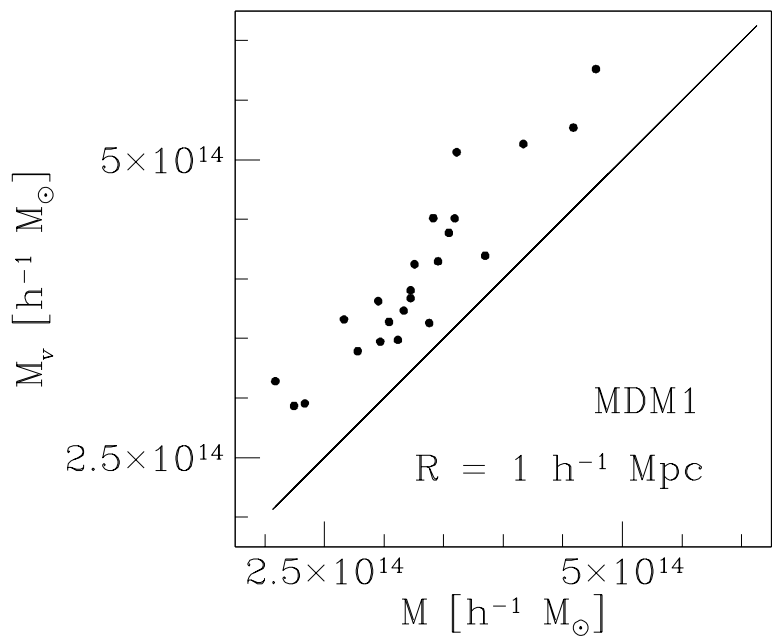
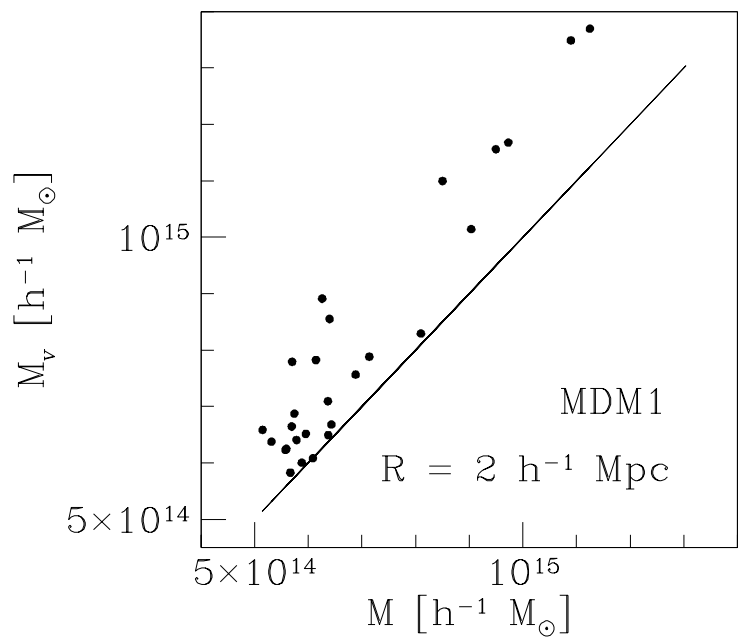
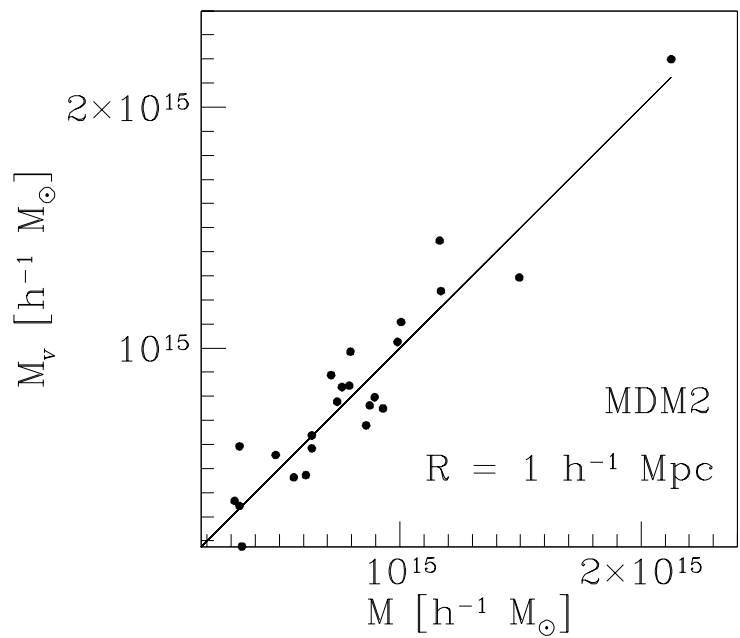
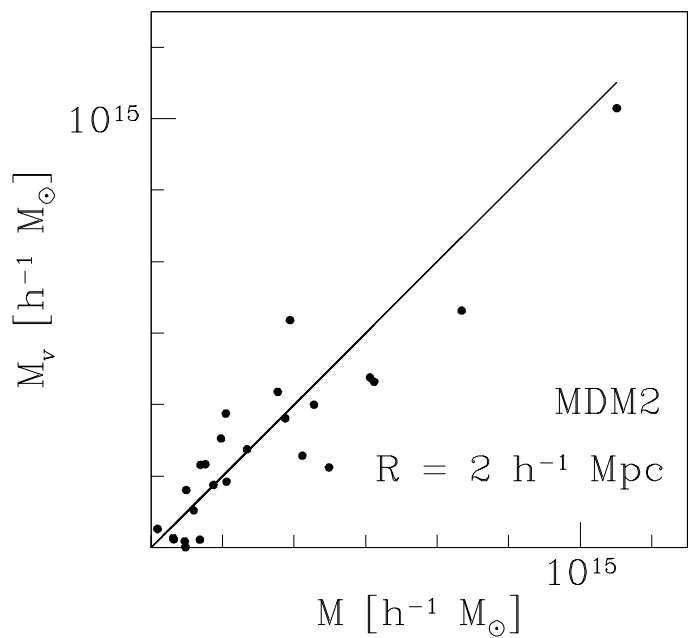
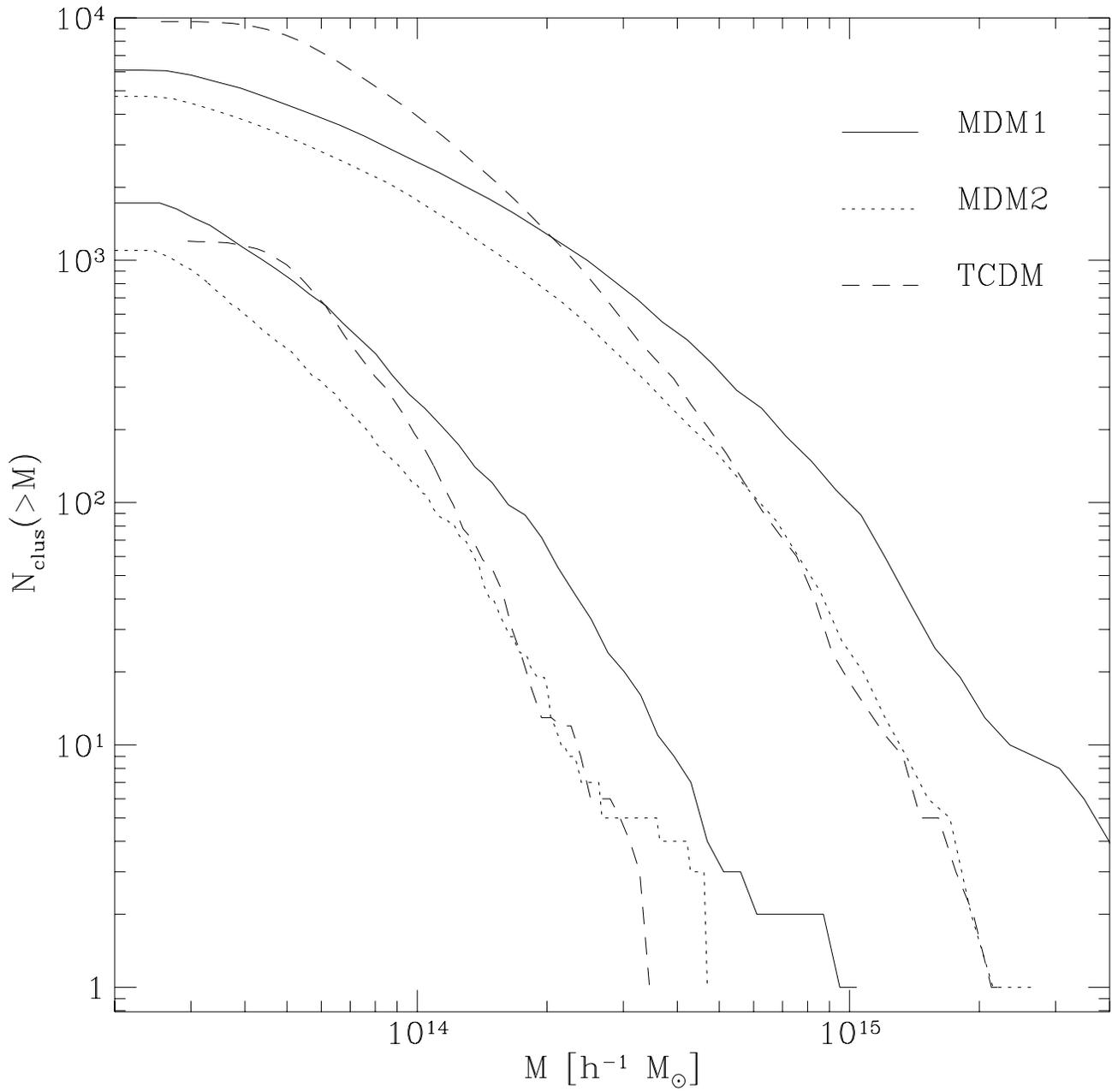
\end{document}